\documentclass[english,tightenlines,eqsecnum,amsmath,amssymb,aps,prd,superscriptaddress,nofootinbib,showpacs,floats]{article}
\usepackage{amsmath,amsfonts,amssymb,multicol}
\usepackage{graphicx,caption,subcaption}
\usepackage{epstopdf}
\usepackage[usenames]{xcolor}
\usepackage{epstopdf}
\definecolor{AHZ}{rgb}{0.0,0.9,0.2}
\usepackage[colorlinks,linkcolor=AHZ,citecolor=red,bookmarks]{hyperref}
\usepackage{rotating}
\usepackage[mathscr]{eucal}
\usepackage{graphicx}
\usepackage[T1]{fontenc}
\usepackage{babel}
\usepackage{authblk}
\usepackage{epsfig}
\usepackage{color}
\usepackage{amssymb}
\usepackage{fancyhdr}

\def\nn{\nonumber\\}
\newcommand{\f}[2]{\frac{#1}{#2}}
\def\be{\begin{equation}}
\def\ee{\end{equation}}
\def\bea{\begin{eqnarray}}
\def\eea{\end{eqnarray}}

\def\bwt{\begin{widetext}}
\def\ewt{\end{widetext}}

\usepackage{amsmath}

\usepackage{amssymb}

\begin{document}

\title{Gravitational Collapse without Singularity Formation in Brans-Dicke Theory}
\author[1]{A. H. Ziaie\thanks{ah.ziaie@riaam.ac.ir}}
\author[2]{H. Shabani\thanks{H.Shabani@phys.usb.ac.ir}}
\author[1]{H. Moradpour\thanks{h.moradpour@riaam.ac.ir}}
\affil[1]{Research~Institute~for~Astronomy~and~Astrophysics~of~ Maragha~(RIAAM), University of Maragheh,  P.~O.~Box~55136-553,~Maragheh, Iran}
\affil[2]{Physics Department, Faculty of Sciences, University of Sistan and Baluchestan, Zahedan, Iran}
\renewcommand\Authands{ and }
\maketitle
\begin{abstract}
In the present work we study collapse process of a homogeneous and isotropic fluid in Brans-Dicke ({BD}) theory with non-vanishing spacetime torsion. In this theory, torsion can be generated by the {BD} scalar field as well as the intrinsic angular momentum (spin) of matter. Assuming the matter content of the collapsing body to be a Weyssenhoff fluid, which is a generalization of perfect fluid in general relativity ({GR}) in order to include the spin effects, we find that in BD theory with torsion, the existence of spin effects could avoid the spacetime singularity that forms in the original version of this theory~\cite{saulbhd}. Numerical simulations of collapse model show that the spacetime singularity is replaced by a non-singular bounce, the spacetime event at which the collapse process halts at a minimum radius and then turns into an expanding phase. Moreover, the model parameters can be set so that the apparent horizon will never meet the boundary of the collapsing body so that the bounce event can be detectable by external observers in the Universe.
\end{abstract}
\section{Introduction}
The process of gravitational collapse in GR has been studied for many years since the pioneering work by Oppenheimer, Snyder, and Datt (OSD)~\cite{OSD} showed that collapse of a spherical matter distribution under its own gravity leads to the formation of a black hole (BH). In this simple study where the collapsing matter is described by a homogeneous pressure-less fluid, a spacetime singularity forms as the collapse end state which is necessarily hidden by an event horizon and therefore invisible to far away observers. Since then, several models have been considered in the literature with the aim of understanding possible extensions of OSD collapse scenario. In the context of GR, singularity theorems predict that under physical circumstances (requirements for the matter content i.e., energy conditions) the final fate of gravitational collapse scenario would be the formation of a spacetime singularity, an event at which densities as well as curvatures grow unboundedly and become infinite~\cite{12}. It is generally believed that in these extreme spacetime events the classical framework of GR theory breaks down~\cite{3t,4mal}. Unlike the BH case, further investigations showed that for certain matter distributions that satisfy physically reasonable conditions the spacetime singularity that arises as the collapse final outcome is not hidden behind a horizon and thus can causally connect to far away observers, see, e.g.,~\cite{joshmal} and references therein. These spacetime events are called naked singularities that the existence of which has been also reported in modified gravity theories~\cite{nsmod}. Although GR has proven to be very successful  theory in describing gravitational phenomena, the occurrence of spacetime singularities signal that this theory is utilized beyond its domain of validity. Therefore, it is widely believed that in the late stages of a collapse scenario where super dense regimes of extreme gravity are present, GR must be replaced by a suitable modified gravity theory in order to cure the singularity problem. This is because the formation of a spacetime singularity can cause issues such as future predictability and path incompleteness~\cite{futpath}. During the past decades, much effort has gone into searching for physically reasonable collapse settings being free of spacetime singularities. The outcomes of research indicate that in comparison to singular spacetimes in GR, the presence of additional/correction terms within the classical framework of modified gravity theories can lead to singularity avoidance. Such non-singular scenarios have been reported in $f(R)$ gravity~\cite{frbamba} where it is shown that adding higher order contributions of Ricci scalar could cure the curvature singularity in the collapse process. In the context of Einstein-Cartan (EC) theory~\cite{ECNON,JaHaZia} it is shown that ingredients mimicking spin effects (associated with fermions) can replace the spacetime singularity by a non-singular bounce. Also, in this theory, nonsingular cosmological models have been studied in~\cite{ecnonsin} and in~\cite{bambicol} it is shown that a non-minimal coupling between gravity and fermions can influence the endstate of a classical collapse scenario long before quantum gravity effects become important. 
\par
Nevertheless, it is a general belief that in very late stages of a collapse process where super-ultra-dense regions of extreme gravity are present, the Planck scale physics become dominant and it is the nontrivial quantum gravity effects that finally remove the classical singularity~\cite{QuanGra,QuanGra1}. Much attempts have been made so far in this direction in order to find a consistent model which includes quantum effects in a collapse setting. In this regard, though a full quantum theory of gravity has not yet been developed, concerted efforts have been made to take into account the quantum aspects of gravity in the regions near the singularity in terms of semi-classical effects, either by quantizing the matter sector in the field equations of GR or by quantizing certain limited degrees of freedom of the spacetime metric~\cite{4mal}. In~\cite{joshicoll}, quantum effects in gravitational collapse of a scalar field which classically ends in formation of a naked singularity has been considered and it is shown that these effects lead to the dissolution of the collapsing cloud before the singularity can form. Singularity removal in inhomogeneous scalar field collapse with quantum gravity corrections has been presented in~\cite{Husain}. The possibility of singularity avoidance in a homogeneous collapse process with modified matter source inspired by the quantum gravity effects has been discussed in~\cite{malafa}. See also~\cite{otherqgc} for other nonsingular collapse scenarios and~\cite{cosmolqg} for nonsingular quantum cosmological models.
\par 
Among the modified gravity theories, the BD theory is one of the simplest and well studied generalizations of GR where the gravitational coupling constant is replaced by a varying scalar field~\cite{BDTpp}. This theory in which, gravitational interactions are described by two fundamental non-matter fields, i.e., the spacetime metric $g_{\mu\nu}$ and a non-minimally coupled scalar field $\phi$, is an attempt towards incorporating Mach's principle into GR framework~\cite{BDTpp}. The BD scalar field, in the Jordan representation~\cite{jorrep}, couples to gravity with an adjustable parameter, $\omega$, therefore acting as a mediator between matter fields and spacetime geometry. Since its advent, the {BD} theory has attracted attention of many researchers as it arises naturally from supergravity models, string theories at low energies and dimensional reduction of Kaluza-Klein theories~\cite{BDTpp2}. In addition to its cosmological implications~\cite{jorrep},\cite{bdcosmol}, in the past years, the issue of gravitational collapse and its final outcome in BD theory has been studied comprehensively under various models among which we can quote: collapse of an ideal gas in BD theory~\cite{idealg}, investigation of the waveform and amplitude of scalar-type gravitational waves in spherically symmetric dust fluid collapse~\cite{stwave}, perfect fluid as well as null fluid collapse and formation of naked singularity~\cite{nsbd,rudranull}, collapse of a self-gravitating electrically charged scalar field~\cite{sfelcha}, dynamics of string-inspired charged black holes~\cite{stringbdbh} and critical behavior in BD collapse~\cite{critbd}. Numerical simulation of Oppenheimer-Snyder collapse in BD theory has been performed in~\cite{saulbhd} where the authors concluded that black holes that arise as the end product of the collapse scenario are identical to those of GR in final equilibrium, but are completely different from those of GR during dynamical evolution.
\par
Recently, a modified version of {BD} theory with non-vanishing torsion has been investigated in~\cite{BDTTOR1986} and it has been shown that in addition to the spin of matter, the BD scalar field itself has contribution in generating the torsion field. This theory can be viewed as somehow a unification of BD and {EC} theories that we call it hereafter ECBD theory. Cosmological as well as astrophysical aspects of {ECBD} theory has been investigated in~\cite{Sung-Won KIM1987,ECBDCOSMOD} and higher dimensional extension of this theory has been studied in~\cite{HIGHDECBD}. In~\cite{HHECBD}, the authors have studied the existence and stability of Einstein static universe, a cosmological model according to which, our universe admits no timelike singularity, it is ever existing and lives in an almost a static state in the infinite past ($t\rightarrow-\infty$) and then evolves into an inflationary era. In this model, the Universe emerges from the Einstein static state rather than the initial big bang singularity. As it was mentioned earlier, gravitational collapse in BD theory results in formation of spacetime singularities either naked or covered by a horizon. It is therefore well motivated to search for possible nonsingular collapse scenarios that could emerge from ECBD theory in which the spin effects prevent singularity formation. Our paper is then organized as follows: In section~\ref{BDColl} we give a brief review on BD theory and present a singular solution for the collapse process in this theory. In section~\ref{ECBDnColl} we study the evolution of collapse scenarios in {ECBD} theory and seek for nonsingular solutions along with investigating the dynamics of apparent horizon. Sec.~\ref{exte} is devoted to exterior spacetime and we summarize our results in section~\ref{summconc}.
\section{Gravitational Collapse in BD theory}\label{BDColl}
In the present section we give a brief review on collapse scenario in BD theory without considering the effects of spin and torsion. In Jordan frame~\cite{jorrep}, this theory is described by the following action 
\be\label{action}
S=\int d^4x\sqrt{-g}\left[\f{\phi R}{\kappa^2}-\f{\omega}{\phi}g^{\alpha\beta}\nabla_\alpha\phi\nabla_\beta\phi\right]+S^m,
\ee
where $\nabla_\alpha$ denotes covariant derivative with respect to the Christoffel connection $\Gamma^{\alpha}_{\,\,\beta\gamma}$, $S^m$ is the action describing ordinary matter (i.e., any form of physical matter different from the BD scalar field $\phi$), $\omega$ is a dimensionless parameter representing the strength of coupling between the scalar field and the spacetime metric and $\kappa^2=16\pi/c^4$. Varying the above action with respect to $\phi$ and $g_{\mu\nu}$ gives the field equations in BD theory as 
\bea
&&R_{\mu\nu}-\f{1}{2}g_{\mu\nu}R=\f{\kappa^2}{\phi}T_{\mu\nu}+\f{\omega}{\phi^2}\left[\nabla_{\mu}\phi\nabla_{\nu}\phi-\f{1}{2}g_{\mu\nu}\nabla_{\eta}\phi\nabla^{\eta}\phi\right]+\f{1}{\phi}(\nabla_{\mu}\nabla_{\nu}\phi-g_{\mu\nu}\Box\phi),\label{bdfieldeqs}
\\
&&\Box\phi=\f{\kappa^2}{2\omega+3}T^\alpha_\alpha,\label{bdfieldeqs1}
\eea
where $T_{\mu\nu}$ is the energy momentum tensor (EMT) of matter fields which fulfills the conservation equation $T^{\mu\nu}_{;\mu}=0$. We assume an isotropic and homogeneous spacetime for the collapsing body whose matter content obeys a perfect fluid with the EMT $T_{\mu\nu}=pg_{\mu\nu}+(\rho+p)u_\mu u_\nu$, where $\rho$ and $p$ are energy density and pressure of the fluid and $u_\mu$ is its four-vector velocity. A linear equation of state (EoS) $p=w\rho$ is also assumed with $-1\leq w\leq1$. The interior spacetime can be parametrized by a spatially non-flat Friedmann-Lema\^{\i}tre-Robertson-Walker ({FLRW}) line element, given by
\begin{align}\label{metricFRW}
ds^{2}=-dt^{2}+\frac{a^{2}(t)dr^{2}}{1-kr^2}+{\mathcal R}^{2}(t,r)d\Omega^2,
\end{align}
where $k$ denotes the spatial curvature, ${\mathcal R}(t,r)=ra(t)$ is the physical radius of collapsing object and $d\Omega^2$ is the metric on a unit two-sphere. The field equations (\ref{bdfieldeqs}) and (\ref{bdfieldeqs1}) can then be written as (we use the units so that $\kappa^2=1$)
\bea
&&3H^2+\f{3k}{a^2}+3H\f{\dot{\phi}}{\phi}-\f{\omega}{2}\left(\f{\dot{\phi}}{\phi}\right)^2=\f{\rho}{\phi},~~~~H=\f{d{a}/dt}{a}=\f{\dot{a}}{a},\label{fbd1}\\
&&2\dot{H}+3H^2+\f{k}{a^2}+2H\f{\dot{\phi}}{\phi}+\f{\omega}{2}\left(\f{\dot{\phi}}{\phi}\right)^2+\f{\ddot{\phi}}{\phi}=-\f{p}{\phi},\label{fbd2}\\
&&\ddot{\phi}+3H\dot{\phi}=\f{\rho-3p}{2\omega+3},~~~~H=\f{da/dt}{a}=\f{\dot{a}}{a}.\label{fbd3}
\eea
Also, the conservation equation for EMT reads
\be
\f{d\rho}{dt}+3H(\rho+p)=0~~~~\Longrightarrow~~~~\rho(a)=\rho_ia^{-3(1+w)},\label{conseeqs1}
\ee
where $\rho_i$ is the initial energy density at the onset of collapse process. The set of differential equations (\ref{fbd1})-(\ref{fbd3}) are not independent from each other, i.e., differentiating Eq.~(\ref{fbd1}) with respect to time along with substituting for $\ddot{a}$, $\dot{\rho}$ and the term $\dot{a}^2+k$ from Eqs.~(\ref{fbd2}), (\ref{conseeqs1}) and (\ref{fbd1}) leaves us with the evolution equation for BD scalar field, Eq.~(\ref{fbd3}). This means that only the two equations (\ref{fbd2}) and (\ref{fbd3}) are independent. We therefore proceed to find an exact solution for these set of differential equations. To this aim we change the differentiation variable from $t$ to $a$, this gives
\bea
&&2aHH_{,a}+3H^2+\f{k}{a^2}+\f{\rho_i}{\phi}\left(w+\f{1-3w}{2\omega+3}\right)a^{-3(1+w)}-aH^2\f{\phi_{,a}}{\phi}+\f{\omega}{2}a^2H^2\left(\f{\phi_{,a}}{\phi}\right)^2=0,\\\label{fbd2a}
&&\left(4aH^2+a^2HH_{,a}\right)\phi_{,a}+a^2H^2\phi_{,aa}-\rho_i\f{1-3w}{2\omega+3}a^{-3(1+w)}=0.\label{fbd3a}
\eea
Now, taking the functionalities of $H(a)$ and $\phi(a)$ as
\be\label{Hphi}
H(a)=\alpha a^\gamma,~~~~~~\phi(a)=\beta a^\delta,
\ee
an exact solution can be obtained by setting 
\bea\label{alphabeta}
\gamma=-1,~~~~\delta=-(1+3w),~~~~\alpha=-\f{\sqrt{2k}}{\left[(3w-2)w\omega-(\omega+2)\right]^{\f{1}{2}}},~~~~\beta=\f{2+\omega(1+w(2-3w))}{2k(1+3w)(3+2\omega)}\rho_i,
\eea
provided that
\be\label{prov}
\left(-1\leq w<-\frac{1}{3}\land \omega >\frac{2}{3 w^2-2 w-1}\right)\lor \left(-\frac{1}{3}<w<1\land \omega <\frac{2}{3 w^2-2 w-1}\right).
\ee
The above condition guarantees that for positive spatial curvature, $\alpha\in\mathbb{R}^-$. We therefore get the scale factor as
\be\label{sf}
a(t)=a_i+\alpha(t-t_i),~~~t_s=t_i-\f{a_i}{\alpha}.
\ee
The above solution indicates a spacetime singularity that forms at the finite amount of time $t_s$. At this spacetime event, the curvature invariants such as Kretschmann scalar
\be\label{Kretscha}
{\tt K}=\f{12}{a^4}\left[a^2\ddot{a}^2+\dot{a}^4+2k\dot{a}^2+k^2\right],
\ee
and energy density (\ref{conseeqs1}) grow unboundedly and diverge. Hence it is a true curvature singularity that cannot be removed by any coordinate transformation. We note that this singular solution stands for a closed FLRW geometry ($k>0$) and the other singular spacetimes with $k=0$ have been previously addressed in the literature, see e.g.~\cite{nsbd}. Our aim in the next section is to seek for possible non-singular scenarios in modified BD theory in the presence of spacetime torsion.
\section{Non-Singular Collapse Scenario in ECBD theory}\label{ECBDnColl}
Consider a Riemann-Cartan manifold as the background spacetime in which the general connection $\tilde{\Gamma}^\alpha_{\,\beta\mu}$ is defined. The spacetime torsion is given by the antisymmetric part of this connection as
\be\label{tor}
T^\alpha_{\,\mu\nu}=\tilde{\Gamma}^\alpha_{\,\mu\nu}-\tilde{\Gamma}^\alpha_{\,\nu\mu}.
\ee
Assuming that the connection is metric compatible, i.e., $\tilde{\nabla}_\mu g_{\alpha\beta}=0$, where $\tilde{\nabla}_\mu$ is the covariant derivative with respect to $\tilde{\Gamma}^\alpha_{\,\beta\mu}$, we get
\be\label{gencon}
\tilde{\Gamma}^\alpha_{\,\beta\mu}=\Gamma^\alpha_{\,\beta\mu}+K^\alpha_{\,\beta\mu},
\ee
where $\Gamma^\alpha_{\,\beta\mu}$ is the Christoffel connection and $K^\alpha_{\,\beta\mu}$ is the contorsion tensor defined as
\be\label{contt}
K^\alpha_{\,\beta\mu}=\f{1}{2}\left[T^\alpha_{\,\beta\mu}-T^{\,\,\alpha}_{\beta\,\,\mu}-T^{\,\,\beta}_{\alpha\,\,\mu}\right].
\ee
The action in BD theory with torsion is given as~\cite{Sung-Won KIM1987,ECBDCOSMOD}
\be
{\cal S}=\int\sqrt{-g}d^4x\left[\f{\phi\tilde{R}}{\kappa^2}-\f{\omega}{\phi}g_{\alpha\beta}\tilde{\nabla}^\alpha\phi\tilde{\nabla}^\beta\phi+L_m(g_{\alpha\beta},K^\alpha_{\,\beta\gamma},\psi^i)\right],
\ee
where $L_m$ is the Lagrangian for mater field(s) $\psi^i$~\cite{Venzo-2-Hehl}. Similar to EC theory, the Ricci curvature scalar is constructed out of the general connection and is given by
\bea\label{RICCIF}
\tilde{{R}}={R}(\Gamma)+\nabla_{\lambda}{K}^{\lambda\rho}\!\!~_{\rho}-\nabla_{\rho}{K}^{\lambda\rho}\!\!~_{\lambda}+{ K}^{\sigma\mu}\!\!~_{\mu}{K}^{\lambda}\!\!~_{\sigma\lambda}
-{K}^{\sigma\rho}\!\!~_{\nu}{K}^{\nu}\!\!~_{\sigma\rho}.
\eea
We note that in addition to spin degrees of freedom of matter which is a source of spacetime torsion, due to nonminial coupling between BD scalar field and curvature, the BD field itself acts as a source of torsion field. Varying the above action with respect to metric and BD field leaves us with the modified BD field equations
\bea
&-&\!\!\!\!\phi\,{G}_{\alpha\beta}-\f{\phi}{2}\bigg[{K}^{\gamma\,\,\delta}_{\,\,\beta}\,{K}_{\delta\gamma\alpha}+{K}^{\gamma\,\,\delta}_{\,\,\alpha}\,{K}_{\delta\gamma\beta}-{g}_{\alpha\beta}{K}^{\gamma\delta\epsilon}\,{K}_{\delta\epsilon\gamma}+{g}_{\alpha\beta}{K}^{\gamma\delta}_{\,\,\,\,\,\gamma}\,{K}_{\delta\,\,\epsilon}^{\,\,\epsilon}-{K}^{\gamma}_{\,\,\alpha\beta}\,{K}^{\delta}_{\,\,\gamma\delta}\nn&-&\!\!\!\!{K}^{\gamma}_{\,\,\beta\alpha}\,{K}^{\delta}_{\,\,\gamma\delta}\bigg]+\f{1}{2}{K}^{\gamma}_{\,\,\beta\gamma}\nabla_\alpha\phi+\f{1}{2}{K}^{\gamma}_{\,\,\alpha\gamma}\nabla_\beta\phi+\nabla_\alpha\nabla_\beta\phi+\f{\omega}{\phi}\nabla_\alpha\phi\nabla_\beta\phi\nn&+&\!\!\!\!\f{1}{2}\left[{K}^{\gamma}_{\,\,\alpha\beta}-{K}^{\gamma}_{\,\,\beta\alpha}\right]\nabla_\gamma\phi+\f{1}{2}{g}_{\alpha\beta}\left[{K}^{\delta\gamma}_{\,\,\,\,\gamma}-{K}^{\gamma\delta}_{\,\,\,\,\gamma}\right]\nabla_\delta\phi+{g}_{\alpha\beta}\Box\phi\nn&-&\!\!\!\!\f{\omega}{2\phi}{g}_{\alpha\beta}\nabla_\gamma\phi\nabla^\gamma\phi=\f{\kappa^2}{2}{T}_{\alpha\beta},\label{fieldeqsmetric}\\
&-&{K}^{\alpha\beta}_{\,\,\,\,\,\alpha}\,{K}_{\beta\,\,\,\gamma}^{\,\,\,\gamma}+K^{\alpha\beta\gamma}\,{K}_{\beta\gamma\alpha}+{R}-\nabla_\alpha{K}^{\alpha\beta}_{\,\,\,\,\,\beta}+\nabla_\beta{K}^{\alpha\beta}_{\,\,\,\,\,\alpha}+\f{2\omega}{\phi}\Box\phi-\f{\omega}{\phi^2}\nabla_\alpha\phi\nabla^\alpha\phi=0.\label{evoleqbdfiled1}
\eea
where ${T}_{\alpha\beta}=2\left(\delta {L}_m/\delta {g}^{\alpha\beta}\right)/\sqrt{-{g}}$ is the EMT of matter fields. Varying the action with respect to contorsion leads the modified Cartan field equation as
\bea\label{finaleqcontor}
{K}^{\mu}_{\,\,\,\alpha\beta}=-\f{\kappa^2}{4\phi}\left[{S}_\beta^{\,\,\,\mu}{u}_\alpha+{S}_\alpha^{\,\,\mu}{u}_\beta+{S}_{\alpha\beta}{u}^\mu\right]+\f{1}{2\phi}\left[\delta_\beta^{\,\,\mu}\nabla_\alpha\phi-{g}_{\alpha\beta}\nabla^{\mu}\phi\right].
\eea
The quantity $S_{\mu\nu} = -S_{\nu\mu}$ is a second-rank antisymmetric tensor which is defined as the spin density tensor. This quantity in EC theory bears the contribution due to intrinsic angular momentum (spin) of fermionic particles within a perfect fluid~\cite{Venzo-1-Hehl}. Therefore a classical description of spin in EC theory generalizes the perfect fluid of GR to the so-called Weyssenhoff fluid~\cite{KCQG1987W1947}. The EMT on the right hand side of (\ref{fieldeqsmetric}) can be written as sum of the EMT of a usual perfect fluid and an intrinsic spin part, as~\cite{gasprl}
\bea\label{spflutensor}
{T}_{\alpha\beta}=\left[(\rho+p){u}_{\alpha}{u}_{\beta}-p{g}_{\alpha\beta}\right]+\f{\kappa^2}{8\phi}\left[2{S}_{\gamma\delta}\,{S}^{\gamma\delta}{u}_{\alpha}{u}_{\beta}-2{S}_{\alpha}^{\,\,\,\delta}{S}_{\beta\delta}{u}_\gamma{u}^{\gamma}\right]+\f{\nabla^\gamma\phi}{8\phi}\bigg[{S}_{\beta\gamma}{u}_\alpha+{S}_{\alpha\gamma}{u}_\beta\bigg].
\eea
Substituting then for the above EMT into field equations (\ref{fieldeqsmetric}) and (\ref{evoleqbdfiled1}) and after rearranging, we arrive at final version of the field equations in ECBD theory, as
\bea 
{G}_{\alpha\beta}&=&T_{\alpha\beta}^{{\rm sf}}+T_{\alpha\beta}^{\phi}=\f{\kappa^2}{\phi}\left[\left(\rho+p-\f{\kappa^2\sigma^2}{6\phi}\right){u}_{\alpha}{u}_{\beta}-{g}_{\alpha\beta}\left(p-\f{\kappa^2\sigma^2}{12\phi}\right)\right]+\f{3+2\omega}{2\phi^2}\nabla_{\alpha}\phi\nabla_{\beta}\phi\nn&+&\f{1}{\phi}\nabla_{\alpha}\nabla_{\beta}\phi-{g}_{\alpha\beta}\left[\f{\Box\phi}{\phi}+\f{3+2\omega}{4\phi^2}\nabla_{\epsilon}\phi\nabla^{\epsilon}\phi\right],\label{Gmunu}\\
\Box\phi&=&\f{\kappa^2}{2(3+\omega)}\left[\rho-3p-\f{3\kappa^2\sigma^2}{8\phi}\right],\label{eqelolsf}
\eea
where $T_{\alpha\beta}^{{\rm sf}}$ and $T_{\alpha\beta}^{\phi}$ are energy momentum tensors of Weyssenhoff spin fluid and BD scalar field, respectively. The quantity $\sigma^2$ is the square of spin density which is obtained by suitable averaging procedure on spin density tensor~\cite{gasprl}
\bea\label{avgprospinsq}
\langle {S}_{\mu\nu}{S}^{\mu\nu}\rangle=2{\sigma}^2,\nn
\langle{S}_\mu^{\,\,\beta}{S}_{\nu\beta}\rangle=\f{2}{3}\left[{g}_{\mu\nu}-{u}_\mu{u}_\nu\right]{\sigma}^2.
\eea
It is noteworthy that in the process of taking the average of spin density tensor, we assumed a cloud of randomly oriented spin particles so that, the average of the spin density tensor is assumed to be zero, i.e., $\langle {S}^{\mu\nu}\rangle=0$. From Eqs.~(\ref{Gmunu}) and (\ref{eqelolsf}) we observe that the presence of spin effects appear as a negative pressure within the EMT and we expect that such negative pressure term could prevent singularity formation during the dynamical evolution of the collapse scenario. To investigate such possibility we begin by substituting metric (\ref{metricFRW}) into the above field equations, this gives ($\kappa^2=1$)
\bea\label{ecbdfes1}
3\left[H^2+\f{k}{a^2}\right]=\f{1}{\phi}\left[\rho-\f{\sigma^2}{12\phi}\right]+\f{2\omega+3}{4}\f{\dot{\phi}^2}{\phi^2}-3H\f{\dot{\phi}}{\phi},
\eea
\bea\label{ecbdfes2}
-2\f{\ddot{a}}{a}-\f{\dot{a}^2}{a^2}-\f{k}{a^2}=\f{1}{\phi}\left[p-\f{\sigma^2}{12\phi}\right]+2H\f{\dot{\phi}}{\phi}+\f{2\omega+3}{4}\f{\dot{\phi}^2}{\phi^2}+\f{\ddot{\phi}}{\phi},
\eea
\bea\label{evoleqsf0}
\ddot{\phi}+3H\dot{\phi}=\f{1}{2\omega+6}\left[\rho-3p-\f{3\sigma^2}{8\phi}\right].
\eea
We also have conservation equation (\ref{conseeqs1}) for the fluid part and the following relation 
\bea
\f{d\sigma^2}{dt}+6H\sigma^2=0,~~~~\Longrightarrow~~~~\sigma^2(a)=\sigma_i^2a^{-6},\label{conseeqs2}
\eea
for the spin part, where $\sigma_i$ is the initial value for the square of spin density.
\subsection{Approximate solutions}
\begin{figure}
	\begin{center}
		\includegraphics[width=10cm]{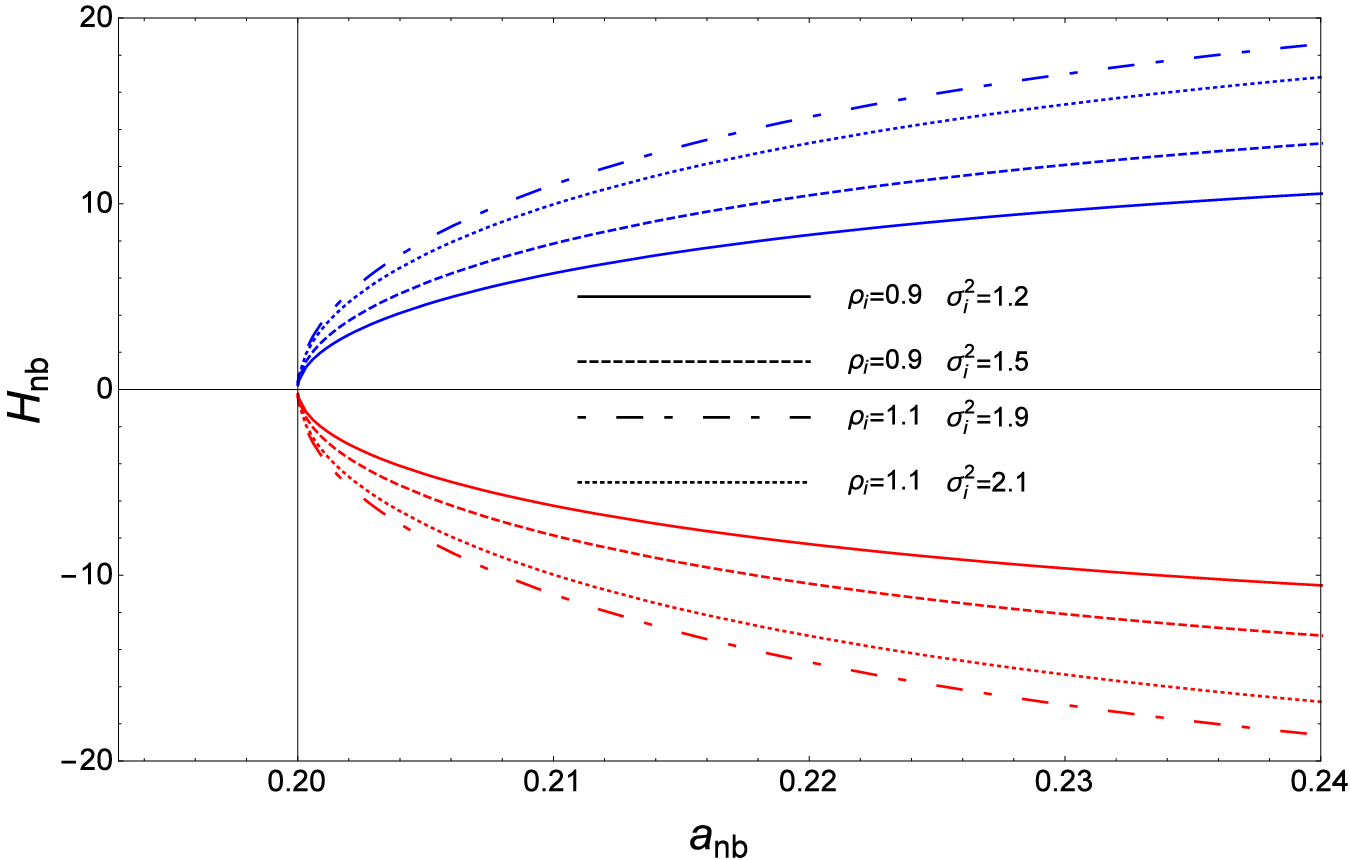}
		\caption{Behavior of the collapse rate against the scale factor in near bounce regime for $w=0.1$, $C_1=1$ and different initial values of energy and square of spin densities. the bounce occurs at $a_b=0.2$ where $H_{nb}(a_b)=0$.}\label{FIG1}
	\end{center}
\end{figure}

We proceed to find a semi-exact solution in the late stages of the collapse scenario where the contracting phase is about to end and then be converted to an expanding one. Similar to the case of BD theory, only equations (\ref{ecbdfes2}) and (\ref{evoleqsf0}) are independent. Using then the change of variable $d/dt=aHd/da$, these two equations can be rewritten as
\bea
2aHH_{,a}\!\!\!\!\!&+&\!\!\!\!\!3H^2+\f{k}{a^2}+\f{\rho_ia^{-3(1+w)}}{\phi}\left[w-\f{1-3w}{2(\omega+3)}\right]-aH^2\f{\phi_{,a}}{\phi}\nn&+&\f{2\omega+3}{4}a^2H^2\left(\f{\phi_{,a}}{\phi}\right)^2-\f{\sigma_i^2}{4a^6\phi^2}\left[\f{1}{3}+\f{3}{4(\omega+3)}\right]=0,\label{twoeqs}\\
\Big(aH^2\!\!\!\!\!&+&\!\!\!\!\!a^2HH_{,a}\Big)\phi_{,a}+a^2H^2\phi_{,aa}+3aH^2\phi_{,a}-\f{\rho_i(1-3w)}{2(\omega+3)\phi}a^{-3(1+w)}+\f{3\sigma_i^2}{16(\omega+3)a^6\phi^2}=0.\label{twoeqs1}
\eea
The above system of differential equations cannot be solved analytically, however, we may employ some approximations in order to simplify it. Consider a collapse scenario that has started its evolution at $t=t_i$ and $a_i=a(t_i)$ and has entered a ``{\it near-bounce}'' regime where the collapse process with the rate $H_{nb}$ is undergoing a decelerated contracting phase and then halts at $a=a_b$, i.e., $H_{nb}(a_b)=0$. In such a situation $H_{nb}^2(a)\ll H_{nb}(a)$ and $a_b<a_{nb}\ll a_i$. Note that $a_{nb}$ is the scale factor at near bounce regime so that $a_{nb}(t_b)=a_b$ where $t_b$ is the time at which the bounce occurs. We therefore neglect the term $k/a^2$ and those containing square of collapse rate. Equations (\ref{twoeqs}) and (\ref{twoeqs1}) then take the form
\bea
2a_{nb}H_{nb}H^\prime_{nb}+\f{\rho_i}{\phi_{nb}}\left(w+\f{3w-1}{2(\omega+3)}\right)a_{nb}^{-3(1+w)}-\f{\sigma_i^2(21+4\omega)}{48a_{nb}^6(\omega+3)\phi_{nb}^2}=0,\label{twoeqapp}\\
a_{nb}^2H_{nb}H^\prime_{nb}\phi^\prime_{nb}-\f{1}{2(\omega+3)}\left[\rho_i(1-3w)a_{nb}^{-3(1+w)}-\f{3\sigma_i^2}{8a_{nb}^6\phi_{nb}}\right]=0,\label{twoeqapp1}
\eea
where a prime denotes differentiation with respect to $a_{nb}$. Now, if we set the BD parameter as $\omega=(1/2w)\left(1-9w\right)$, we obtain an exact solution for the above system as
\bea
\phi_{nb}&=&\left\{C_1[(2+3w)a_{nb}]^{\f{18w}{2+3w}}-\f{16w\rho_ia_{nb}^{3(1-w)}}{(w+2)\sigma_i^2}\right\}^{-1},\label{nbsolphi}\\
H_{nb}&=&\pm\sqrt{2}\left[\f{2+3w}{288(3w-1)}\left(A_1a_{nb}^{-6w}+A_2a_{nb}^{-\f{3(3w^2+w+2)}{2+3w}}+A_3a_{nb}^{6\f{3w-2}{3w+2}}\right)+C_2\right]^{\f{1}{2}},\label{nbsolH}
\eea
where $C_1$ and $C_2$ are constants of integration and
\be\label{A123}
A_1=\f{256w\rho_i^2}{\sigma_i^2(w+2)^2},~~~~A_2=-\f{64C_1w(3w+2)^{\f{21w+2}{3w+2}}\rho_i}{4+(3w+5)w^2},~~~~A_3=-\f{C_1^2\sigma_i^2(3w+2)^{\f{39w+2}{3w+2}}}{3w-2}.
\ee
Next, we proceed to obtain $C_2$ using the condition $H_{nb}(a_b)=0$. This gives
\bea\label{cons2}
C_2&=&\f{3w+2}{288a_b^{6w}}\Bigg[\f{64C_1\rho_iwa_b^{\f{3w^2+9w-2}{3w+2}}(3w+2)^{\f{21w+2}{3w+2}}}{(w+2)(3w-1)(w(3w-1)+2)}-\f{256w\rho_i^2}{(w+2)^2(3w-1)\sigma_i^2}\nn&+&\f{C_1^2\sigma_i^2(3w+2)^{\f{39w+2}{3w+2}}a_b^{6\f{3w^2+9w-2}{3w+2}}}{9w(w-1)+2}\Bigg].
\eea
By substituting the above constant into Eq.~(\ref{nbsolH}) we get the near bounce solution for the collapse rate so that the negative sign stands for contracting regime and the positive one for expanding regime. This behavior has been sketched in Fig.~(\ref{FIG1}), where we observe that $H_{nb}<0$ in the contracting regime (family of red curves) and after the bounce occurs an expanding phase (family of blue curves) with $H_{nb}>0$ commences. In the next section we try to solve Eqs.~(\ref{ecbdfes1})-(\ref{evoleqsf0}) using numerical techniques and find the non-singular collapse solutions under physically reasonable conditions.
\subsection{Numerical solutions}\label{numsols}
\begin{figure}
	\begin{center}
		\includegraphics[width=7.5cm]{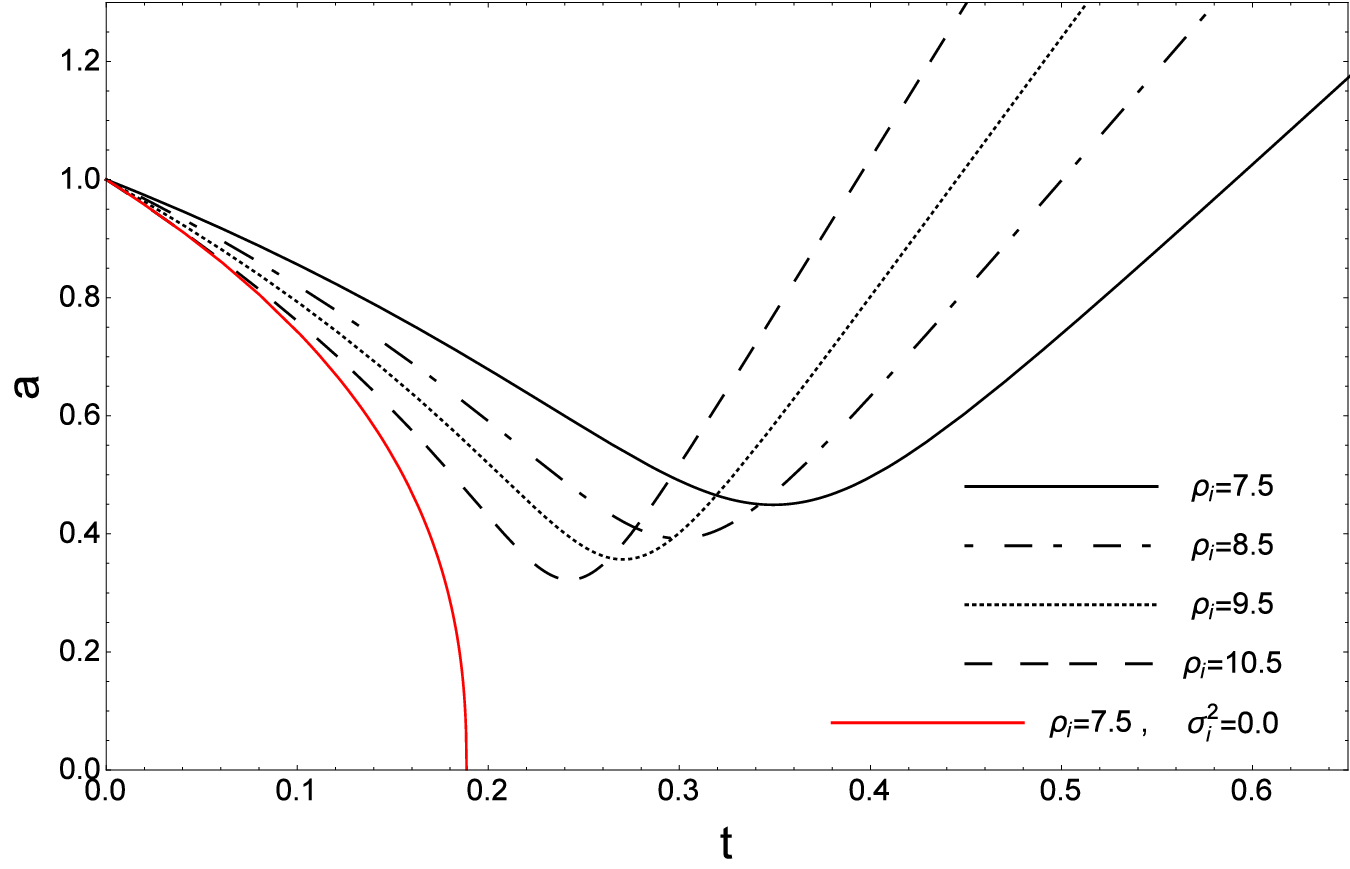}
		\includegraphics[width=7.5cm]{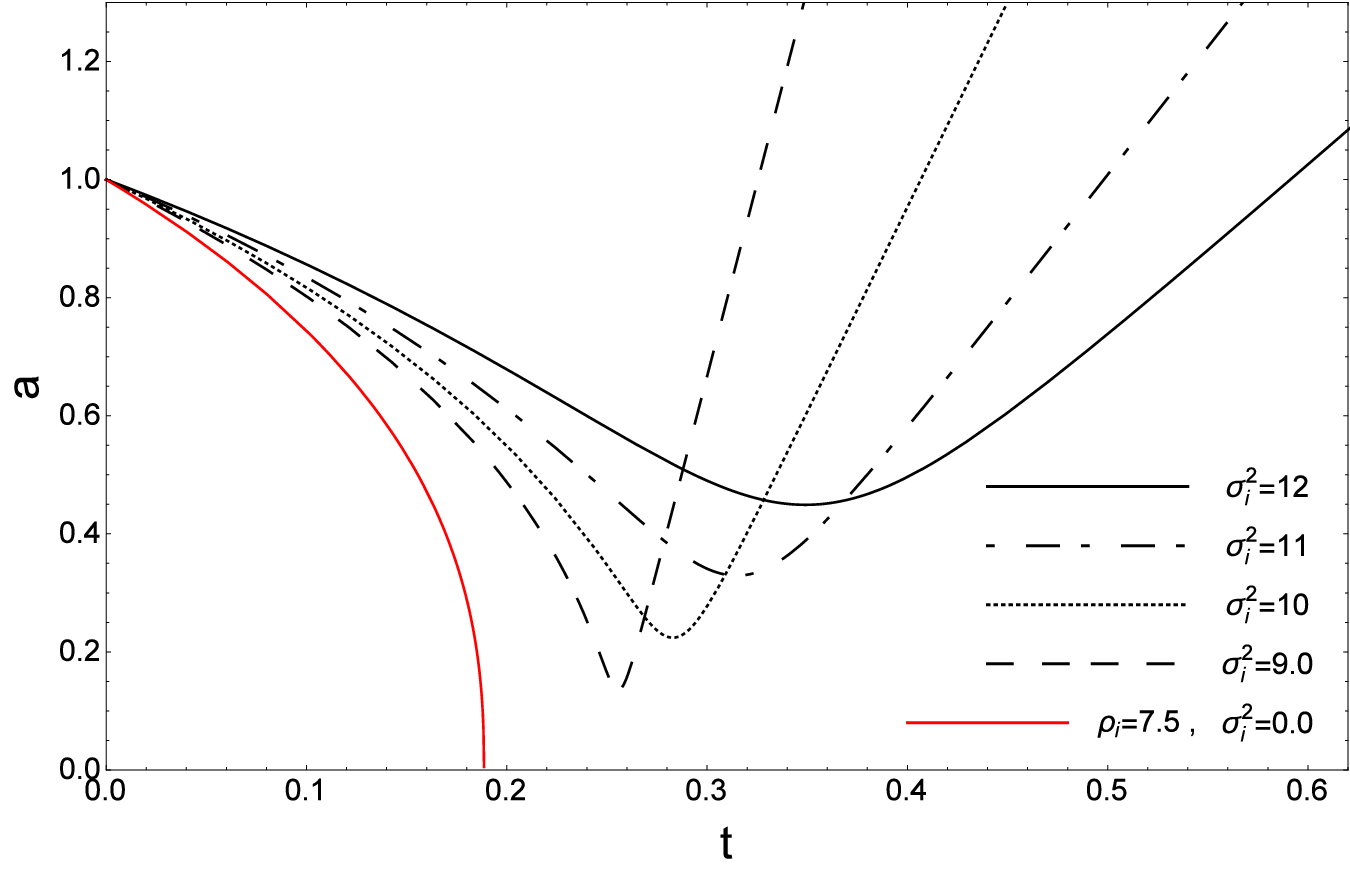}
		\caption{Behavior of scale factor for $k=1$, $w=0$, $\omega=9$, $a_i=1$, $\phi_i=1.5$, $\dot{\phi}_i=1.9$ and different values of initial energy density in the left panel with $\sigma_i^2=12$ and square of spin density in the right panel with $\rho_i=7.5$.}\label{FIG2}
	\end{center}
\end{figure}

\begin{figure}
	\begin{center}
		\includegraphics[width=7.5cm]{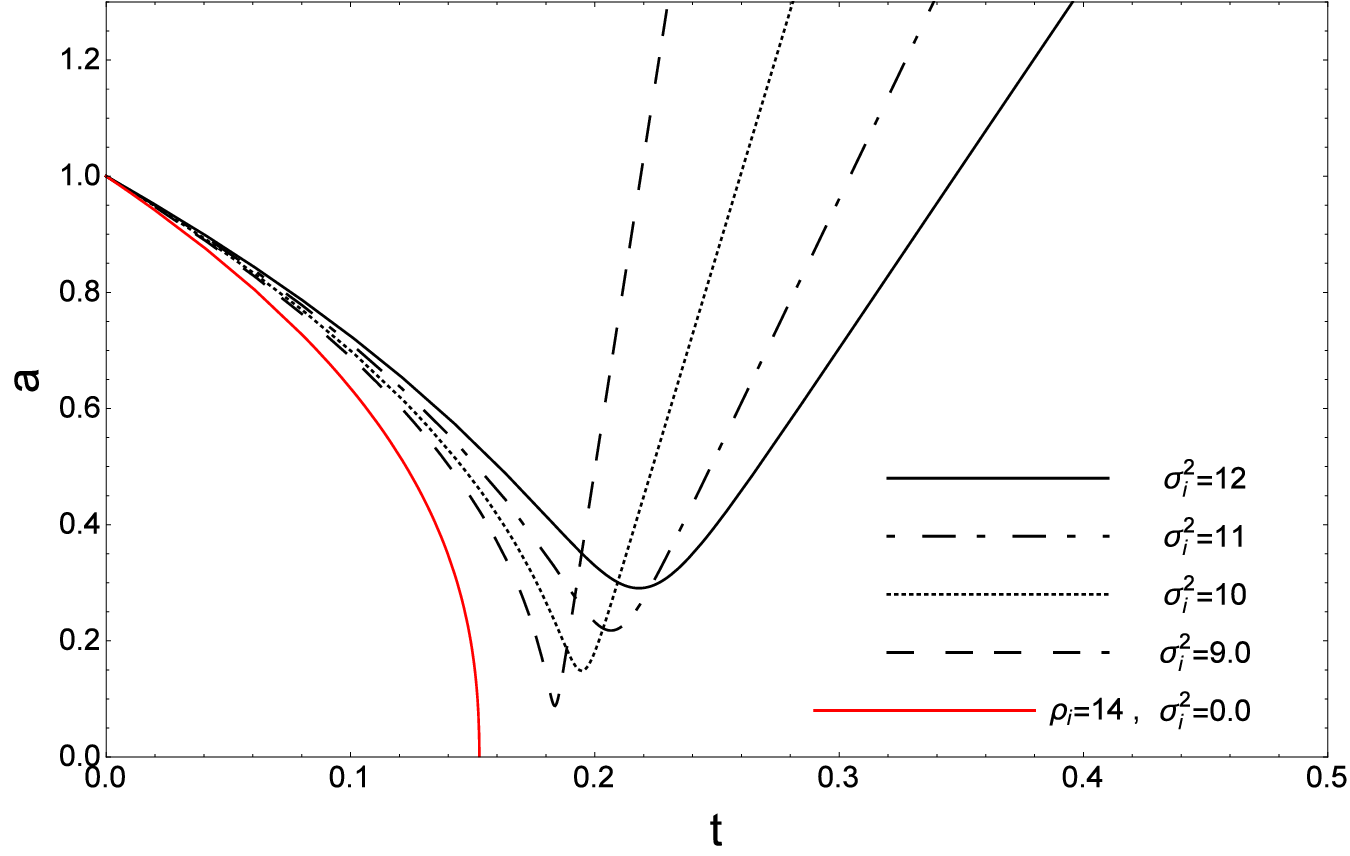}
		\includegraphics[width=7.5cm]{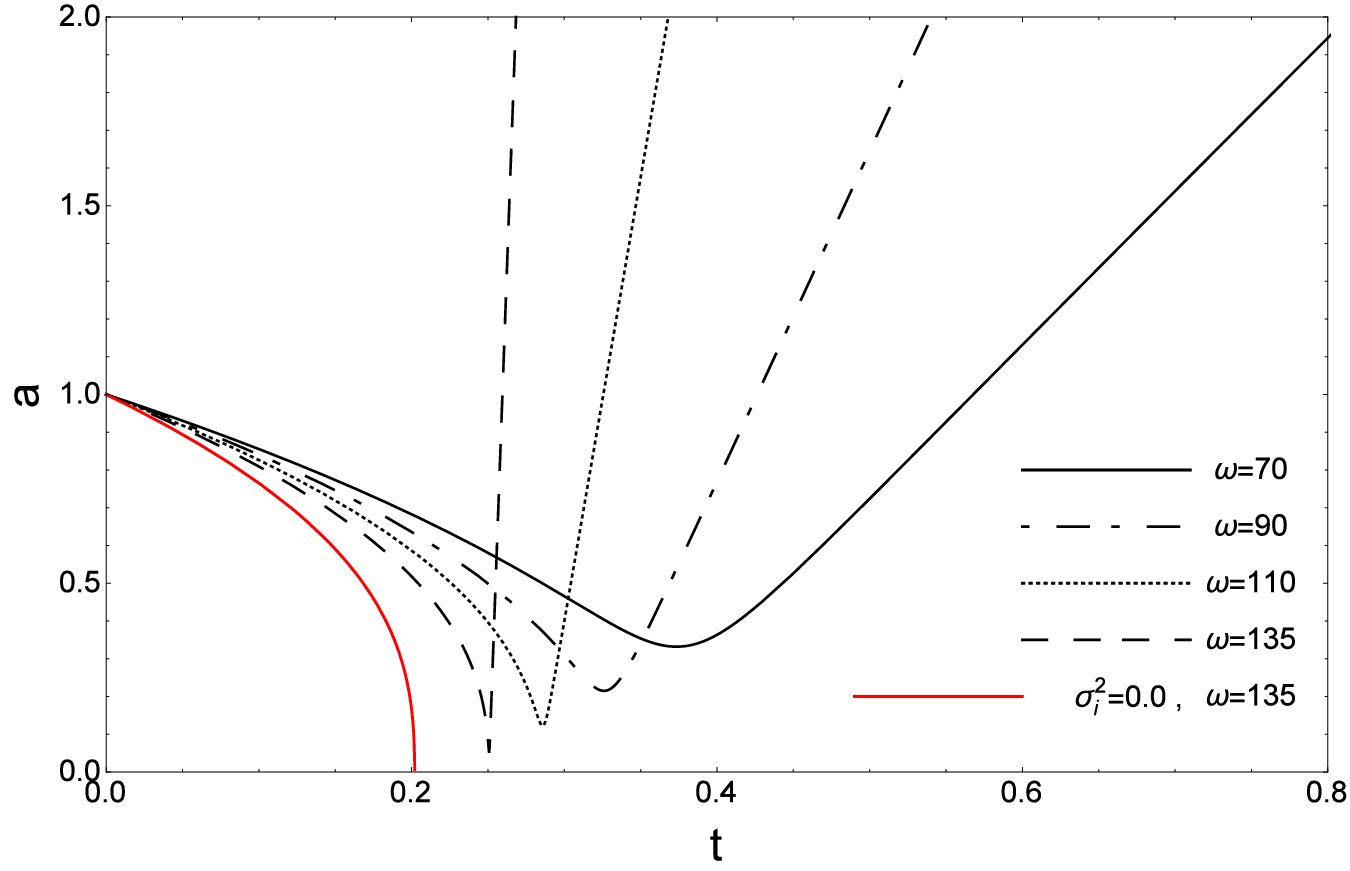}
		\caption{Behavior of scale factor for $k=1$, $w=0$, $\omega=9$, $a_i=1$, $\phi_i=1.5$, $\dot{\phi}_i=1.9$, $\rho_i=14$ and different initial values of square of spin density in the left panel and different values of BD parameter with $\rho_i=10$ and $\sigma_i^2=8$ in the right panel.}\label{FIG3}
	\end{center}
\end{figure}

\begin{figure}
	\begin{center}
		\includegraphics[width=7.5cm]{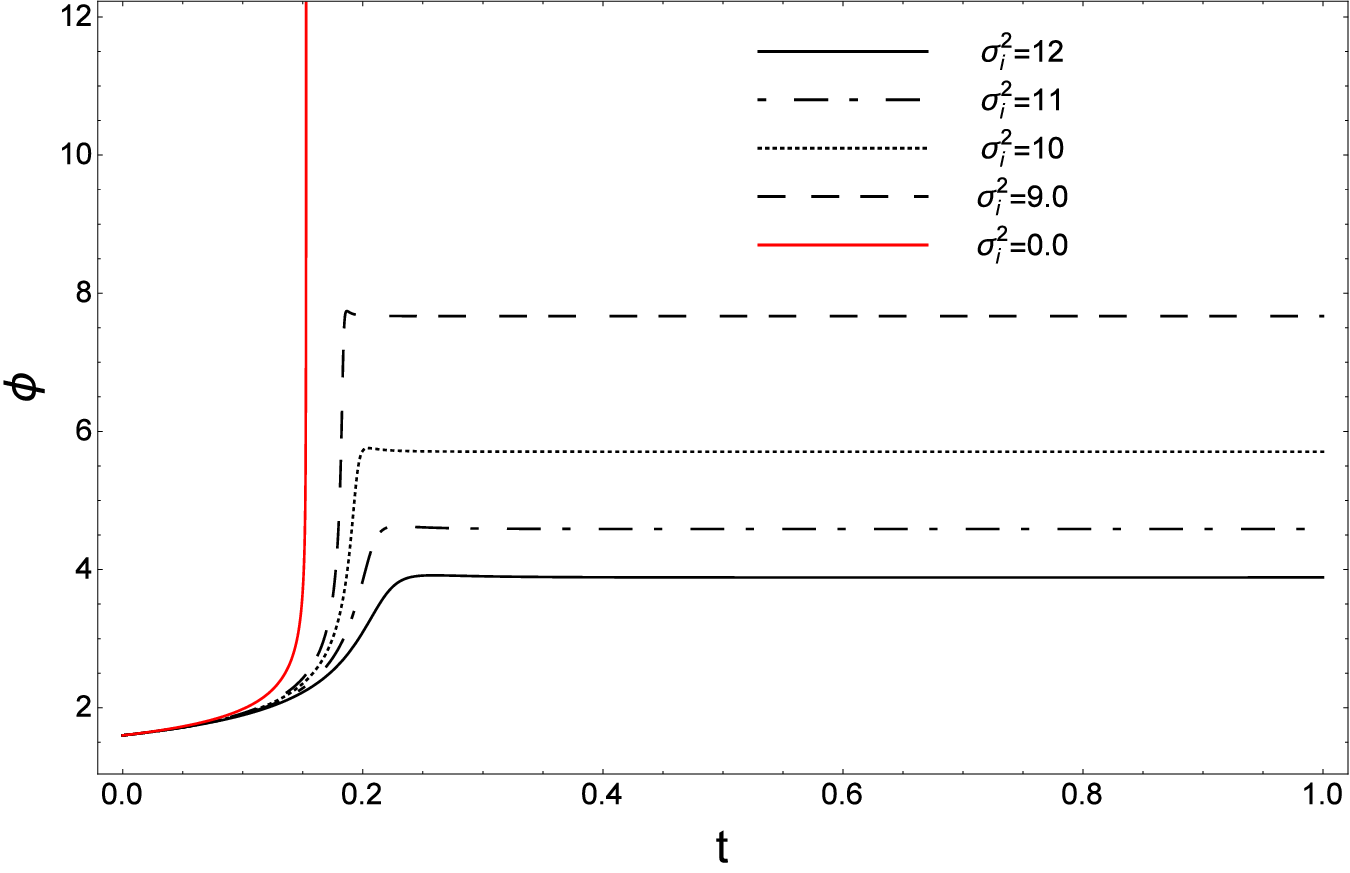}
		\includegraphics[width=7.5cm]{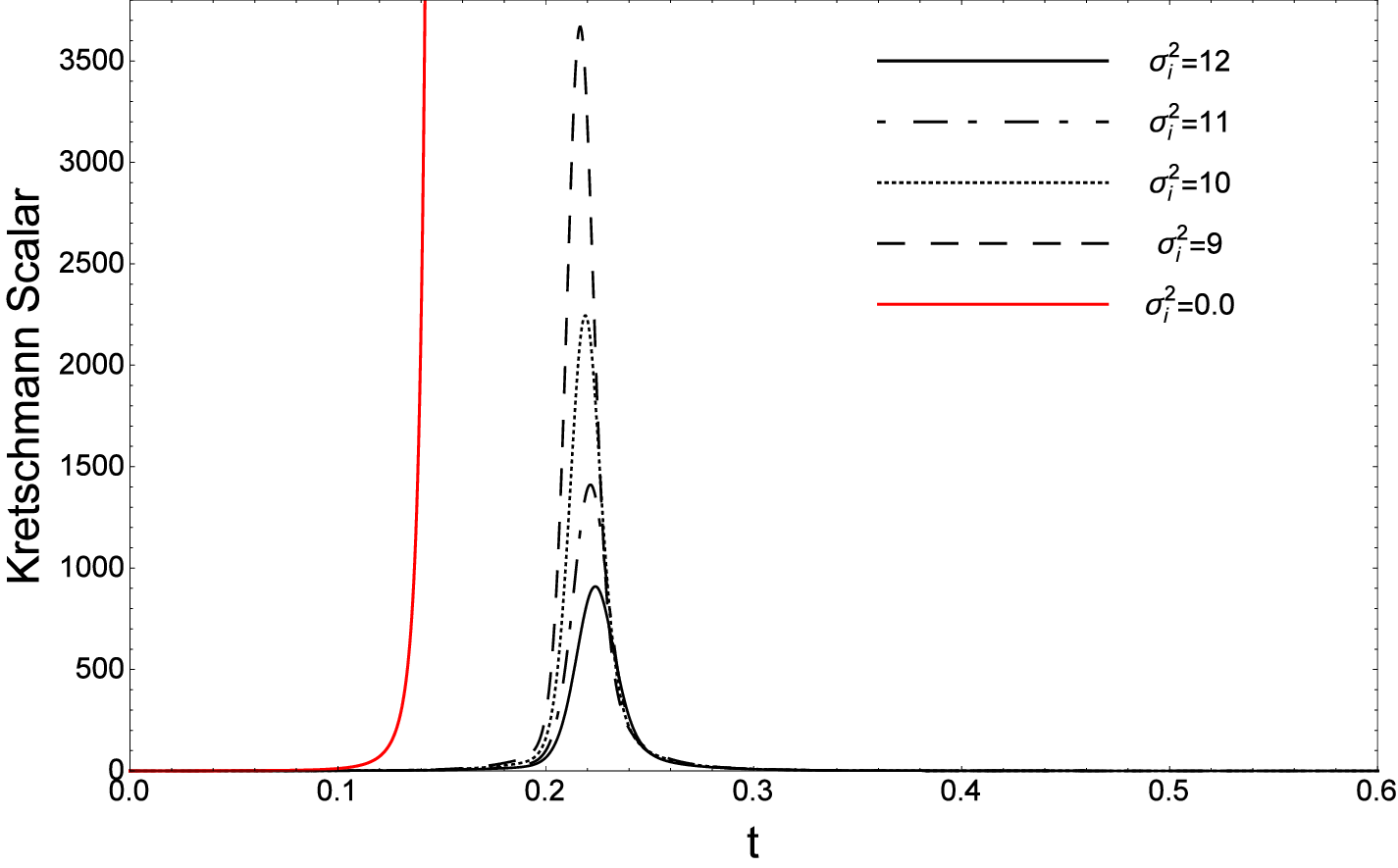}
		\caption{(Left panel) Behavior of BD scalar field for $k=1$, $w=0$, $\omega=9$, $a_i=1$, $\phi_i=1.5$, $\dot{\phi}_i=1.9$, $\rho_i=14$ and different initial values of square of spin density. (Right panel) Time evolution of Kretschmann invariant for the same values of model parameters as of the left panel.}\label{FIG4}
	\end{center}
\end{figure}

\begin{figure}
	\begin{center}
		\includegraphics[width=7.6cm]{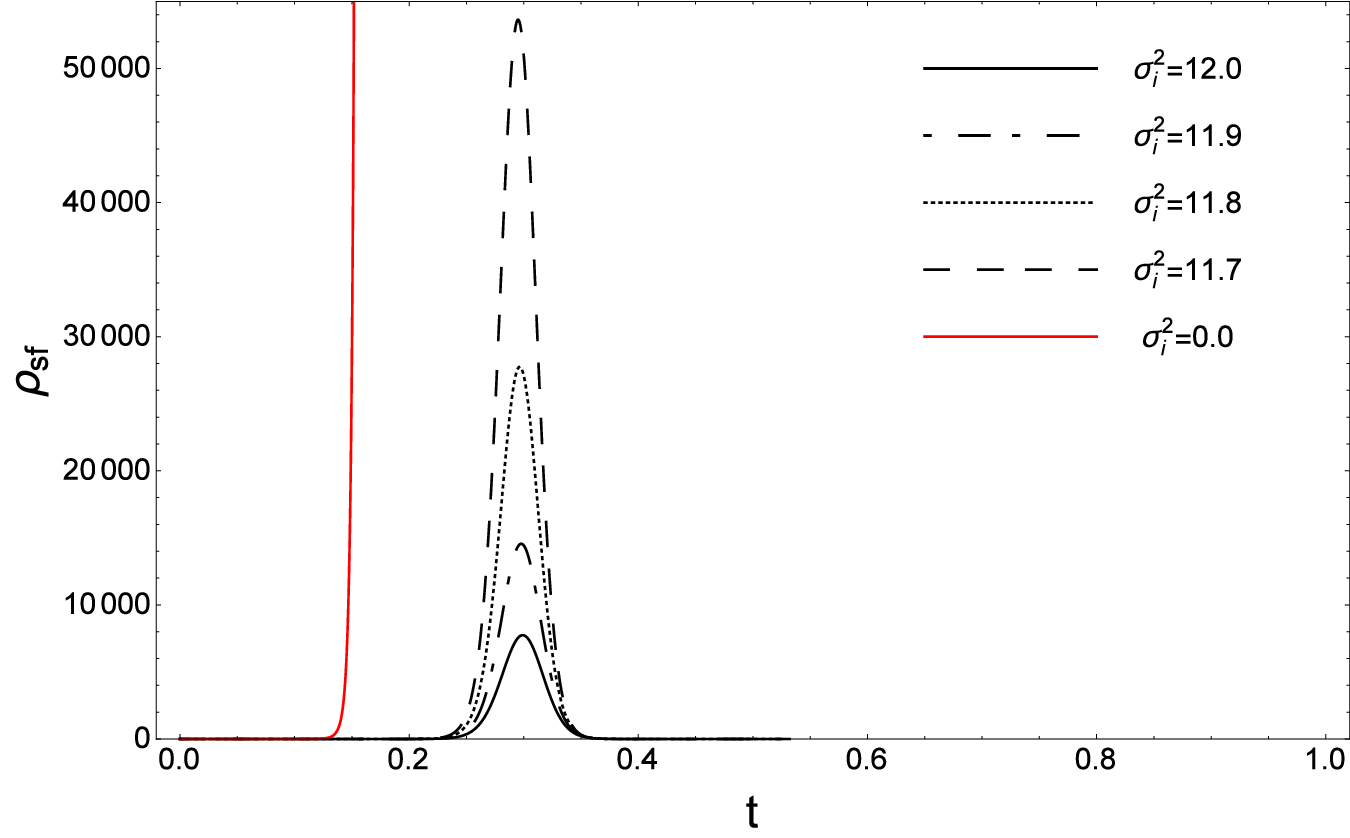}
		\includegraphics[width=7.6cm]{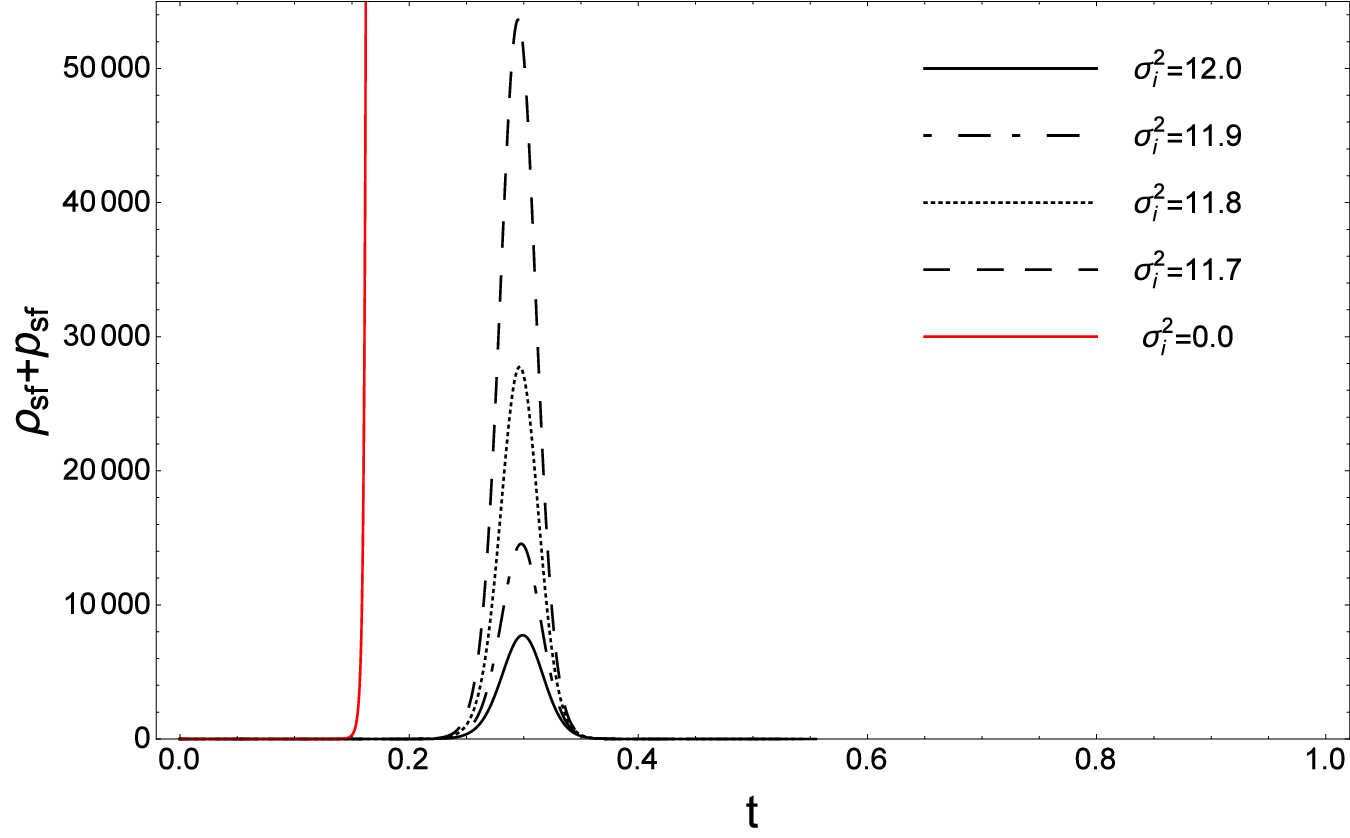}
		\caption{(Left panel): Behavior of energy density of spin fluid for $k=1$, $w=0$, $\omega=9$, $a_i=1$, $\phi_i=1.5$, $\dot{\phi}_i=1.9$, $\rho_i=7.5$ and different initial values of square of spin density. (Right panel) Weak energy condition for spin fluid for the same values of model parameters as of the left panel. The red curve shows the behavior of these quantities when spin effects are not taken into account.}\label{FIG5}
	\end{center}
\end{figure}

\begin{figure}
	\begin{center}
		\includegraphics[width=7.6cm]{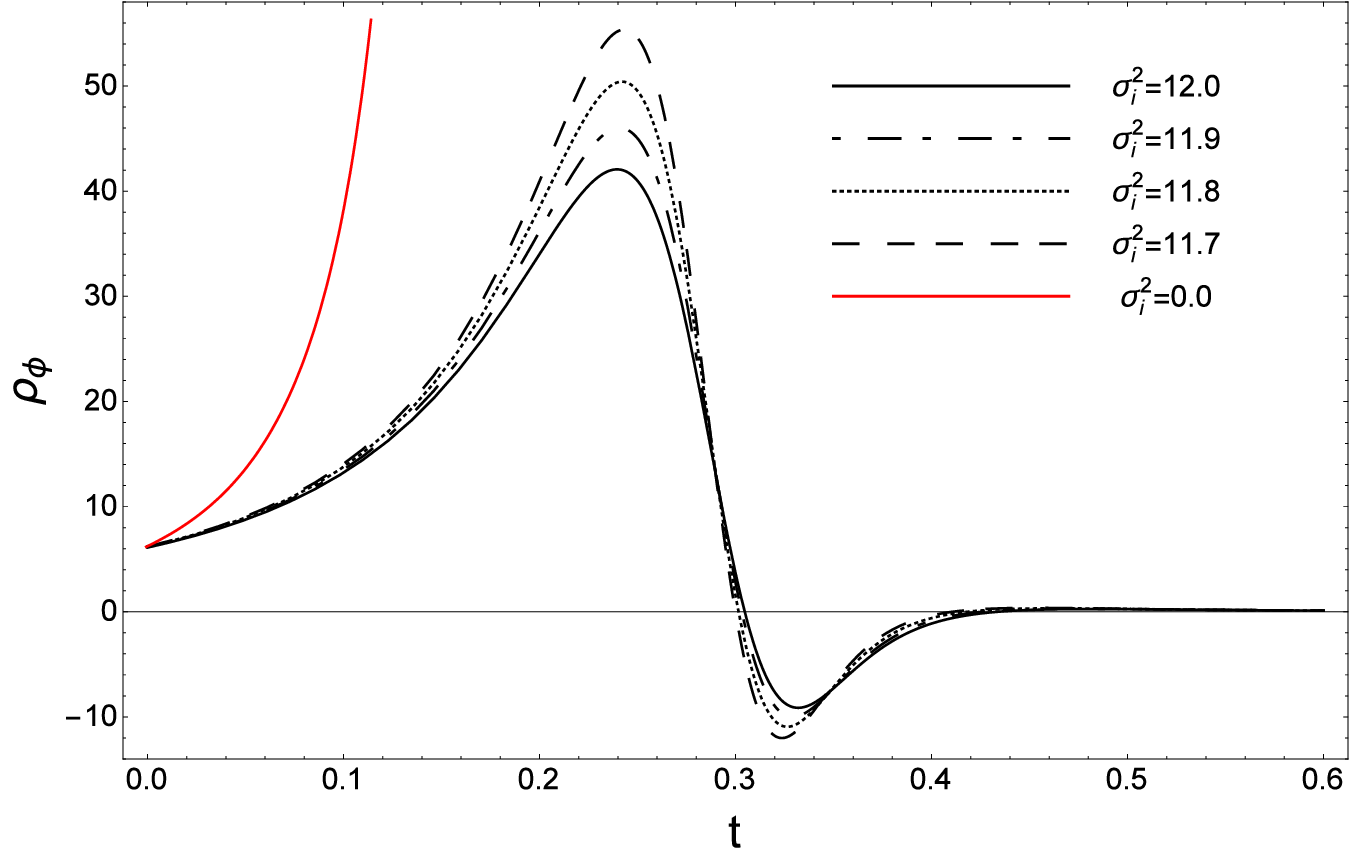}
		\includegraphics[width=7.6cm]{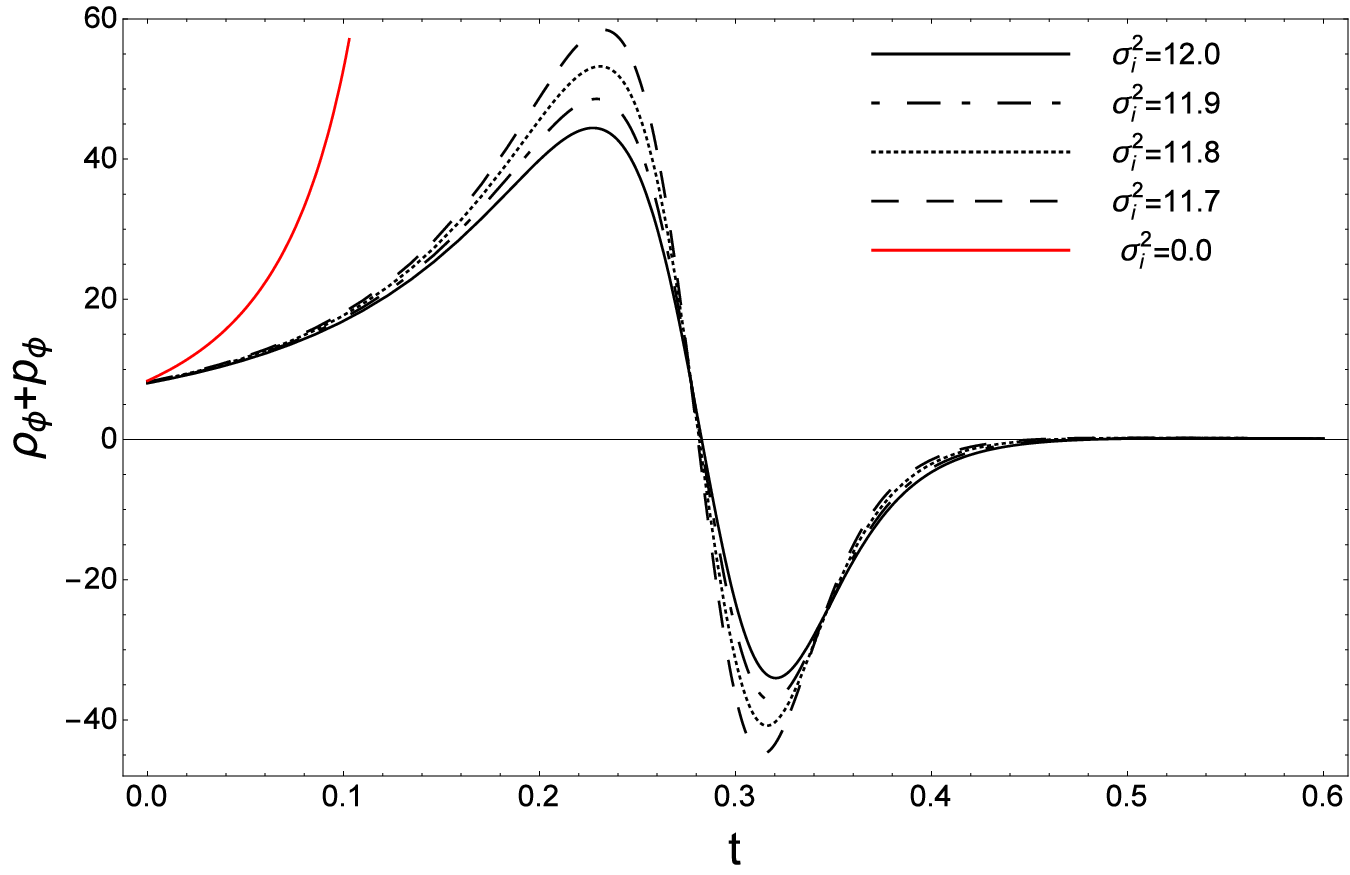}
		\caption{(Left panel): Behavior of energy density of BD scalar field for $k=1$, $w=0$, $\omega=9$, $a_i=1$, $\phi_i=1.5$, $\dot{\phi}_i=1.9$, $\rho_i=7.5$ and different initial values of square of spin density. (Right panel) Weak energy condition for BD scalar field for the same values of model parameters as of the left panel. The red curve shows the behavior of these quantities when spin effects are not taken into account.}\label{FIG6}
	\end{center}
\end{figure}

In the present section we try to solve numerically the system of equations (\ref{ecbdfes1})-(\ref{evoleqsf0}) and estimate the behavior of scale factor and other related quantities over time. Substituting for $\ddot{\phi}$ and $\dot{a}^2$ from Eqs.~(\ref{evoleqsf0}) and (\ref{ecbdfes1}) into Eq.~(\ref{ecbdfes2}) gives
\bea\label{ecbd22}
&&a^{3w}\left[48 (\omega +3)a^5\left(6\phi^2\ddot{a}-6\phi \dot{a}\dot{\phi}+(2\omega+3)a\dot{\phi}^2\right)-\sigma_i^2(16\omega+75)\right]\nn&+&24 \rho _ia^3\phi(w(6\omega+9)+2\omega+9)=0.
\eea
The above equation along with Eq.~(\ref{evoleqsf0}) have to be solved numerically with the initial condition on collapse velocity, $\dot{a}(t_i)=\dot{a}_i$ as provided by Eq.~(\ref{ecbdfes1}). Figure~(\ref{FIG2}) shows the time evolution of scale factor for dust collapse assuming different initial values of energy and square of spin densities. We observe that spin effects provide a completely different evolution for the collapse setting in comparison to the case when these effects are absent. As the family of black curves show, the collapsing body continues its contracting phase until the spin effects become dominant and consequently the collapse process halts at a bounce time where the scale factor reaches a minimum value. After the bounce, the contracting regime is replaced by an expanding one with an increasing scale factor. This non-singular collapse scenario can be compared to the case of singular one, i.e., when the spin effects are not taken into account. In this case the collapsing object continues its contraction until reaching a spacetime singularity, where the scale factor vanishes, see the red curve.  
Comparing the two diagrams in Fig.~(\ref{FIG2}) along with the left panel of Fig.~(\ref{FIG3}), one can deduce that the competition between initial energy and square of spin densities can affect the time at which the bounce occurs and its corresponding scale factor. In the left panel of Fig.~(\ref{FIG2}), for a fixed value of $\sigma_i^2$, increasing $\rho_i$ leads to smaller values of $a_b$ and $t_b$. This scenario also occurs for a fixed value of $\rho_i$ but decreasing $\sigma_i^2$, see the right panel. In other words, if we interpret energy density and square of spin density as agents for attractive nature of gravity and repulsive effects, then initially the more the gravitational attraction or correspondingly the less the repulsive effects, the more the collapsing object needs to be contracted until the spin effects become dominant to finally prevent the singularity formation. The opposite case for this scenario also holds for smaller values of $\rho_i$ and larger ones for $\sigma_i^2$. Likewise, in the left panel of Fig.~(\ref{FIG3}) we observe that even if $\sigma_i^2<\rho_i$, in later stages of the collapse process the repulsive effects due to spin contribution overcome the gravitational pull of gravity to finally avoid the singularity. In the right panel, we have sketched the evolution of scale factor against time for different values of BD parameter. This parameter represents the strength of scalar to tensor coupling to the matter so that the smaller the value of BD parameter, the larger the contribution of the BD scalar field to gravitational interaction. We can therefore deduce that for fixed initial values of energy and spin densities, the BD parameter could affect the softness of the change from a contracting regime to an expanding one. In this regard, we may conclude that the stronger the contribution of the BD scalar field (which here also acts as a source of dynamical torsion) to the gravitational interaction (i.e., the smaller values of $\omega$), the softer the bounce occurs. The left panel in Fig.~(\ref{FIG4}) shows the time evolution of BD scalar field where we observe that in comparison to the case of vanishing spin effects (red curve), the BD field is finite and assumes a bounded maximum value at the bounce. This diagram can be compared to the left panel of Fig.~(\ref{FIG3}) in such a way that, the lower the initial value for square of spin density, the higher this maximum for BD field would be and the more the collapsing cloud needs to be contracted until the gravity succumbs to the repulsive spin effects. The right panel in Fig.~(\ref{FIG4}) shows the behavior of Kretschmann scalar over time where we observe that this quantity behaves regularly throughout the collapse process and reaches a maximum finite value at the bounce. It then converges to zero in the post-bounce regime where the cloud disperses. However, in the absence of spin effects, this quantity grows unboundedly and diverges at the spacetime singularity, see the red curve. For the sake of physical validity of the collapse model, it is required that the collapse process obeys the weak energy condition (WEC), i.e., $T_{\alpha\beta}v^\alpha v^\beta\geq0$ for all non-spacelike vector fields. This condition means that the energy density as measured by any local observer is non-negative. For EMT given in Eq.~(\ref{Gmunu}) the WEC leads to the following inequalities
\be\label{weak}
\rho_{{\rm sf}}\geq0,~~~~\rho_{{\rm sf}}+p_{{\rm sf}}\geq0,~~~~~~~~\rho_\phi\geq0,~~~~\rho_\phi+p_\phi\geq0,
\ee
or in terms of EMT components, using Eqs.~(\ref{ecbdfes1}) and (\ref{ecbdfes2})
\be\label{weak1}
\f{\rho}{\phi}-\f{\sigma^2}{12\phi^2}\geq0,~~~~(1+w)\f{\rho}{\phi}-\f{\sigma^2}{6\phi^2}\geq0,~~~\f{2\omega+3}{4}\f{\dot{\phi}^2}{\phi^2}-3H\f{\dot{\phi}}{\phi}\geq0,~~~~\f{2\omega+3}{2}\f{\dot{\phi}^2}{\phi^2}-H\f{\dot{\phi}}{\phi}+\f{\ddot{\phi}}{\phi}\geq0.
\ee
In Fig.~(\ref{FIG5}) we have plotted the behavior of WEC over time where in the left panel we observe that the energy density of spin fluid is finite (family of black curves) and positive hence the first inequality of WEC for spin fluid in Eq.~(\ref{weak}) is satisfied. As is seen in the right panel, the second inequality also holds throughout the evolution of collapsing body. It should be noted that though the growth rate of square of spin density is three times bigger (for the dust case) than that of matter density, it is the presence of $\phi^2$ term in the denominator of square of spin density that balances the growth rate of densities and thus preserving the WEC. Figure (\ref{FIG6}) indicates the behavior of energy density profile for BD field and the corresponding WEC, i.e., the last part of Eq.~(\ref{weak}). It is seen that $\rho_\phi$ reaches a maximum positive value at the bounce point and then converges to negative values in the post bounce regime where the collapse rate changes its sign from contracting phase to an expanding one. It then tends to zero at later times. This behavior also holds for $\rho_\phi+p_\phi\geq0$. Thus, the WEC for scalar field profile is violated in the post bounce regime, however for the effective EMT, as given by right hand side of Eqs.~(\ref{ecbdfes1}) and (\ref{ecbdfes2}), the WEC is fulfilled.

\par
We also note that when spin effects are absent i.e., $\sigma_i^2=0$, the ECBD field equations (\ref{ecbdfes1})-(\ref{evoleqsf0}) will be reduced to their BD counterparts\footnote{The geometry is still non-Riemannian as from Eq.~(\ref{finaleqcontor}) the contorsion tensor is nonzero.}, i.e., Eqs.~(\ref{fbd1})-(\ref{fbd3}), if we take the BD parameter as $\omega\rightarrow\omega-3/2$. However, the collapse setting arising from this case is singular as we observe from the red curves in Figs.~(\ref{FIG2})-(\ref{FIG4}), hence, it is the mere presence of repulsive spin effects that could avoid singularity formation in BD theory. Finally, in the limit where $\omega\rightarrow\infty$ and $\phi\rightarrow {\rm const.}$, the ECBD theory reduces to EC theory in which a nonsingular collapse model has been studied earlier~\cite{JaHaZia}.
\subsection{Dynamics of apparent horizon}\label{DAH}
\begin{figure}
	\begin{center}
		\includegraphics[width=7.6cm]{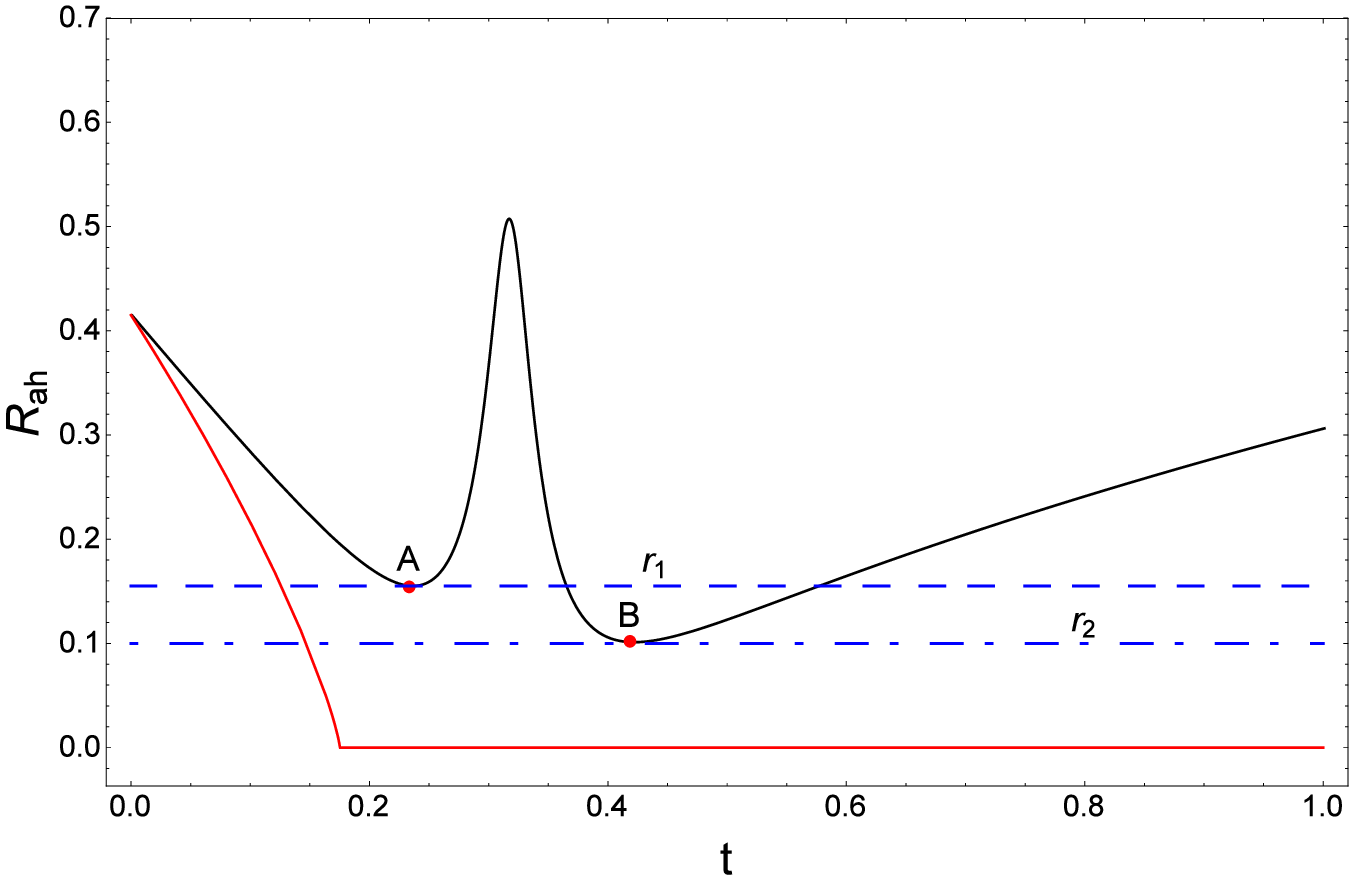}
		\includegraphics[width=7.6cm]{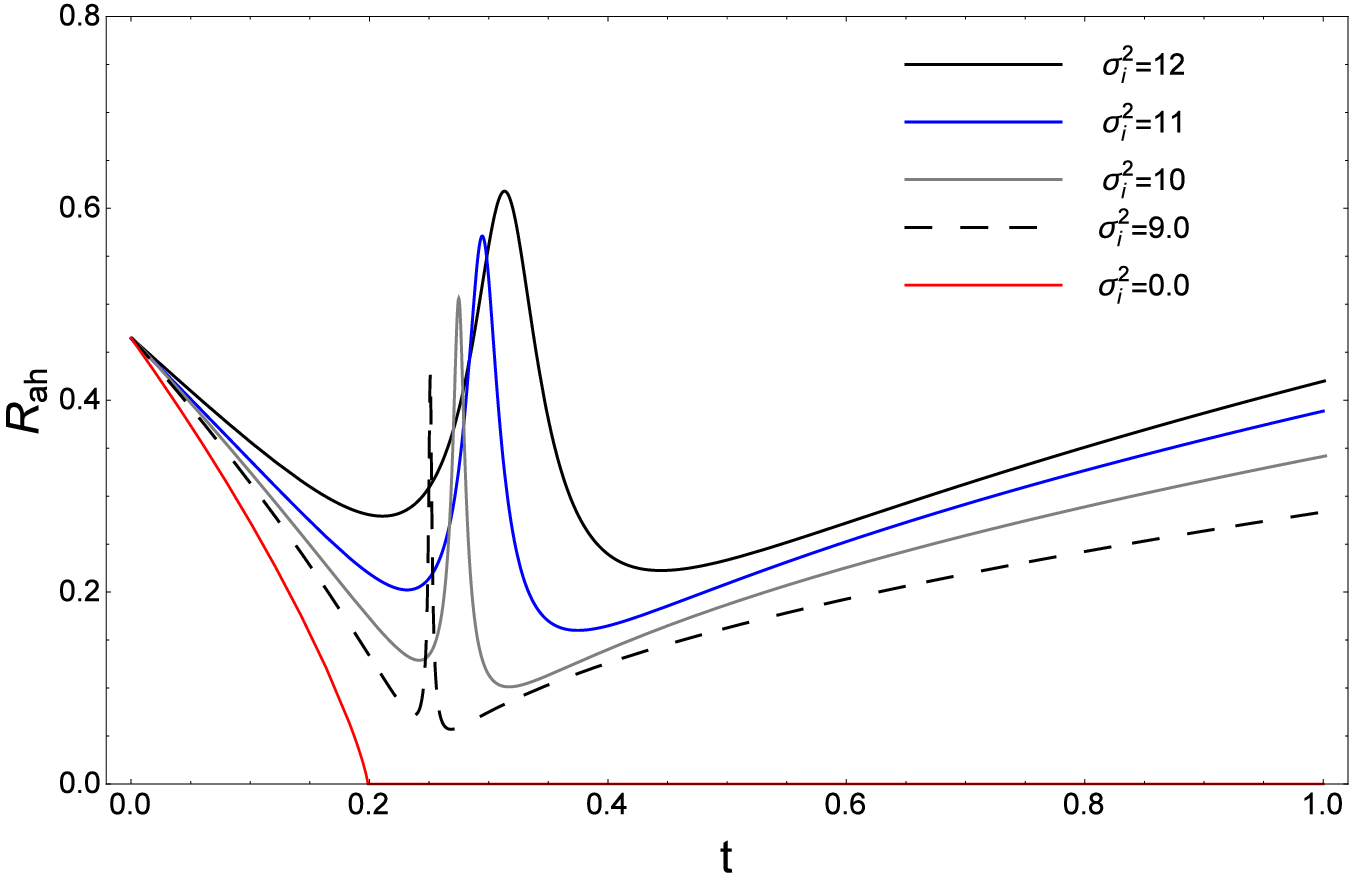}
		\caption{Evolution of apparent horizon curve for $k=1$, $w=0$, $\omega=9$, $a_i=1$, $\phi_i=1.5$, $\dot{\phi}_i=1.9$, $\rho_i=7.5$, $\sigma_i^2=12$ (left panel) and different values of square of spin density (right panel).}\label{FIG7}
	\end{center}
\end{figure}

An important point that needs to be investigated in every collapse setting is the behavior of apparent horizon which is the outermost boundary of the trapped region. Indeed, the apparent horizon determines the boundary between null trajectories that have light-cones confined in the trapped region and thus are causally disconnected from the rest of the universe, and those trajectories that can propagate to far away observers. The condition for the formation of apparent horizon is given by the requirement that the surface ${\mathcal R}={\rm const.}$ is null, i.e., $g^{\mu\nu}\partial_\mu {\mathcal R}\partial_\nu {\mathcal R}=0$. Hence, a shell labeled by the comoving radial coordinate $r$ will become trapped if $\dot{{\mathcal R}}_{ah}^2(t,r)=1$~\cite{QuanGra}. For the line element (\ref{metricFRW}) this condition gives the physical areal radius of apparent horizon as
\be\label{rah}
{\mathcal R}_{ah}(t,r)=\f{a}{\sqrt{1+\dot{a}^2}}.
\ee
The above equation describes time dependence of apparent horizon curve which indeed implies that depending on the behavior of the collapse velocity, the apparent horizon could either form or be avoided. This depends on how the initial radius of the collapsing cloud, i.e., ${\mathcal R}(t_i, r_\Sigma) = r_\Sigma$, is chosen. Fig.~(\ref{FIG7}) shows the dynamics of apparent horizon where in the left panel, we observe that in the singular case, the apparent horizon is a decreasing function of time and as $t\rightarrow t_s$ the areal radius of apparent horizon tends to zero to finally cover the spacetime singularity. In the case of non-vanishing spin effects, the apparent horizon decreases for a while in the contracting regime reaching its first local minimum (Point A) and then increases to a maximum value at the bounce point. It then converges to the next nonzero local minimum (Point B) in the expanding phase and monotonically grows during the post-bounce regime. We therefore find that the dynamics of apparent horizon is completely different when spin effects are present and depending on initial radius of the collapsing body it can be totally avoided. More precisely, if we choose the boundary of the cloud at onset of collapse as $r_\Sigma=r_1$, then as the collapse proceeds, the collapse velocity and the scale factor will achieve the values $\left\{\dot{a}_1,a_1\right\}$ at point A and consequently the apparent horizon equation (\ref{rah}) will be satisfied. This means that a dynamical horizon would form in the contracting regime to meet the boundary of the collapsing object. For $r_2<r_\Sigma<r_1$, the horizon equation will be fulfilled two times in the expanding regime, however, the contracting regime will be free of horizons. For $r_\Sigma=r_2$ the horizon equation will be satisfied once, i.e., when the collapse velocity and the scale factor reach the values $\left\{\dot{a}_2,a_2\right\}$ at point B and hence only one horizon will form to intersect the boundary of the cloud in the expanding phase. Finally for $r_\Sigma<r_2$, we have no horizon formation throughout the collapsing and expanding regimes and thus, the bounce can be visible to an external observer. We can therefore conclude that, unlike the singular case, it is the boundedness of the collapse velocity that, depending on suitable values of $r_\Sigma$, allows the avoidance of horizon formation during the dynamical evolution of the collapsing object. In the right panel we have sketched the time behavior of apparent horizon curve for different values of square of spin density. It is therefore seen that the lower the initial value of square of spin density, the smaller the initial radius of the collapsing body must be taken in order to avoid horizon formation. 
\section{Exterior spacetime}\label{exte}
The collapse scenario we have studied so far deals with the interior spacetime of the contracting object. In order to complete the collapse model we need to match the interior spacetime to a suitable exterior region whose boundary $r=r_{\Sigma}$ is the surface of the collapsing matter. Let us take $({\rm V}^{\pm},g^{\pm})$ as four dimensional spacetimes and $\Sigma$ as a timelike three-dimensional smooth hypersurface separating spacetime into two regions ${\rm V}^+$ (exterior) and ${\rm V}^-$ (interior) with scalar fields $\phi^+$ and $\phi^-$, respectively. The hypersurface $\Sigma$ then results from isometric pasting of two hypersurfaces $\Sigma^{+}$ and $\Sigma^{-}$, which respectively, bound the four-dimensional exterior (${\rm V}^{+}$) and interior (${\rm V}^{-}$) spacetimes. For the interior region we take the line element (\ref{metricFRW}) in the FLRW form as
\be \label{int}
ds_{-}^2=-dt^2+\f{a^2(t)}{1-kr^2}dr^2+a^2(t)(r^2d\theta^2+r^2\sin^2\theta d\varphi^2),
\ee
where the interior coordinates are labeled as $\left\{x_{-}^{\mu}\right\}\equiv\left\{t,r,\theta,\varphi\right\}$. Utilizing generalized Israel-Darmois junction conditions~\cite{gidjcbd} we proceed to match the above line element to the exterior one through the boundary surface $r=r_{\Sigma}$. The line element for the exterior region is chosen as generalized Vaidya metric \cite{vaidyaext} which in retarded (exploding) null coordinates reads
\be \label{out}
ds_{+}^2=-f(v,{\mathtt R})dv^2-2dvd{\mathtt R}+{\mathtt R}^2(d\theta^2+\sin^2\theta d\varphi^2),
\ee
where $f(v,{\mathtt R})=1-2{\mathtt M(v,\mathtt R)}/{\mathtt R}$ with ${\mathtt M(v,\mathtt R)}$ being the Vaidya mass and the exterior coordinates are labeled as $\left\{x_{\rm +}^{\mu}\right\}\equiv\left\{v,{\mathtt R},\theta,\varphi\right\}$. The intrinsic line element on hypersurface $\Sigma$ is given by~\cite{toolkit}
\be\label{intrinsicmetric}
ds_{\Sigma}^2=-d\tau^2+Y^2(\tau)(d\theta^2+\sin^2\theta d\varphi^2),
\ee
where $y^a=\left\{\tau,\theta,\varphi\right\}$, $(a=0,2,3)$ are the intrinsic coordinates of $\Sigma$ with $\tau$ being the time coordinate defined on it and we have assumed the angular coordinates $\theta$ and $\varphi$ are continuous across $\Sigma$. The governing equations of hypersurface $\Sigma$ in the coordinates $x_{\pm}^{\mu}$ are given by
\be\label{eqsigma}
r-r_{\Sigma}=0~~~~~\text{in}~{\rm V}^+,~~~~~~{\mathtt R}-{\mathtt R}_{\Sigma}=0~~~~~ \text{in}~ {\rm V}^-,
\ee
with the help of which we get the interior and exterior induced metrics on $\Sigma^{-}$ and $\Sigma^{+}$, respectively as
\be\label{INMIN}
ds_{\Sigma^{-}}^2=-dt^2+a^2(t)r_{\Sigma}^2(d\theta^2+\sin^2\theta d\varphi^2),
\ee
and
\bea \label{INMEX}
ds_{\Sigma^{+}}^2=-\left[f\big(v(t),{\mathtt R}(t)\big)\dot{v}^2+2\dot{{\mathtt R}}\dot{v}\right]dt^2
+{\mathtt R}^2(t)(d\theta^2+\sin^2\theta d\varphi^2).
\eea
The junction conditions for the first fundamental forms i.e., induced metrics, then give
\be\label{FFF}
ds_{\Sigma}^2=ds_{\Sigma^{-}}^2=ds_{\Sigma^{+}}^2,
\ee
whence we get 
\be\label{FFF1}
f\big(v(t),{\mathtt R}(t)\big)\dot{v}^2+2\dot{{\mathtt R}}\dot{v}=1,~~~~~~{\mathtt R}(t)=r_{\Sigma}a(t)=Y(\tau),
\ee
where an overdot denotes $d/dt$. Next, we proceed to calculate the extrinsic curvature tensors of $\Sigma^+$ and $\Sigma^-$. To this aim we need unit vector fields normal to these hypersurfaces which are given by
\bea\label{NVF}
n^{-}_{\mu}=\left[0,\f{a(t)}{\sqrt{1-r^2}},0,0\right],~~~ n^{+}_{\mu}=\frac{1}{\left[f(v,{\mathtt R})\dot{v}^2+2\dot{{\mathtt R}}\dot{v}\right]^{\frac{1}{2}}}\left[-\dot{\mathtt R},\dot{v},0,0\right].
\eea
Let us take $x^{\mu}=x^{\mu}(y^a)$ as the parametric equations of the hypersurface $\Sigma$. The extrinsic curvature or second fundamental form of this hypersurface is a three-tensor defined as~\cite{toolkit}
\be\label{EXCDEFINE}
{\mathcal K}_{ab}\equiv e_{a}^{\nu}e_{b}^{\mu}\nabla_{\mu}n_{\nu},
\ee
where $e_{a}^{\nu}=\partial x^{\nu}/\partial y^a$ are the basis vectors tangent to the hypersurafce $\Sigma$ and the covariant derivative is taken with respect to the Christoffel $\Gamma^\gamma_{\,\nu\mu}$ connection.
The extrinsic curvature tensors associated with $\Sigma^\pm$ then read
\be\label{EXC}
{\mathcal K}^{\pm}_{ab}=-n_{\mu}^{\pm}\left[\frac{\partial^2x_{\pm}^{\mu}}{\partial y^a\partial y^b}+\tilde{\Gamma}^{\mu\pm}_{\nu\sigma}\frac{\partial x_{\pm}^{\nu}}{\partial y^a}\f{\partial x_{\pm}^{\sigma}}{\partial y^b}\right].
\ee
where $x_{+}^{\mu}(y^{a})$ and $x_{-}^{\mu}(y^{a})$ are parametric relations for the hypersurfaces $\Sigma^{+}$ and $\Sigma^{-}$ and $\tilde{\Gamma}^{\mu\pm}_{\nu\sigma}$ are the components of connections associated to interior and exterior line elements. It should be noted that, to calculate the extrinsic curvature components for the interior spacetime we should take the connection as given by Eq.~(\ref{gencon}) while for the exterior spacetime one may use the usual Christoffel connection. 
\par
From Eqs.~(\ref{Gmunu}) and (\ref{eqelolsf}) we get the following expression for Ricci tensor
\bea\label{Ricc}
R_{\alpha\beta}=\f{\kappa^2}{\phi}\left[T_{\alpha\beta}^{{\rm sf}}-\f{2\omega+5}{4(\omega+3)}g_{\alpha\beta}T^{{\rm sf}}\right]+\f{3+2\omega}{2\phi^2}\nabla_{\alpha}\phi\nabla_{\beta}\phi+\f{1}{\phi}\nabla_{\alpha}\nabla_{\beta}\phi,
\eea
with the help of which the junction conditions take the form \cite{gidjcbd}
\bea
\left[{\mathcal K}^a\,\!\!_b\right]&=&\f{1}{\phi}\left(\Pi^a\,\!\!_b-\f{2\omega+5}{4(\omega+3)}\Pi\delta^a\,\!\!_b\right),~~~~~\left[{\mathcal K}\right]=-\f{(2\omega+3)\Pi}{4(\omega+3)\phi},\label{juncmbd}\\
\left[\phi_{,n}\right]&=&\f{\Pi}{2(\omega+3)},~~~~\left[\phi\right]=0,\label{juncmbd1}
\eea
where the notation $\left[\Psi\right]=\Psi^+|_\Sigma-\Psi^-|_\Sigma$ stands for the jump of given field across the hypersurface $\Sigma$, $n$ labels the coordinate normal to this surface and $\Pi_{ab}={\rm diag}(\rho_{\rm s},p_{\rm s},p_{\rm s})$ is the stress-energy tensor of matter fields (except the BD scalar field) on the shell located at $\Sigma$. The quantities ${\mathcal K}$ and $\Pi=2p_{\rm s}-\rho_{\rm s}$ are traces of ${\mathcal K}^a\,\!\!_b$ and $\Pi^a\,\!\!_b$ with $\rho_{\rm s}$ and $p_{\rm s}$ being the surface energy density and surface pressure, respectively. We also note that from equation (\ref{juncmbd}) we can find the surface stress-energy as
\be\label{eq39equiv}
\Pi^a\,\!\!_b={\phi}\left(\left[{\mathcal K}^a\,\!\!_b\right]-\f{2\omega+5}{2\omega+3}\left[{\mathcal K}\right]\delta^a\,\!\!_b\right).
\ee
The non-vanishing components of Christoffel connection for the exterior spacetime are found as
\bea\label{CHREX}
\Gamma^{v+}_{vv}=-\frac{ f_{,{\mathtt R}}}{2},~~~\Gamma^{{\mathtt R}+}_{vv}=\frac{1}{2}\left( f_{,v}+ff_{,{\mathtt R}}\right),~~~\Gamma^{{\mathtt R}+}_{v{\mathtt R}}=\frac{ f_{,{\mathtt R}}}{2},
\eea
where $,{\mathtt R}\equiv\partial/\partial{\mathtt R}$ and $,v\equiv\partial/\partial v$. Substituting the above expressions into equation (\ref{EXC}) along with using the second part of (\ref{NVF}) we get
\bea\label{EXCEX}
{\mathcal K}^{+t}_{t}=\frac{\dot{v}^2\left[ff_{,{\mathtt R}}\dot{v}+f_{,v}\dot{v}+3f_{,{\mathtt R}}\dot{\mathtt R}\right]+2\left(\dot{v}\ddot{{\mathtt R}}-\dot{{\mathtt R}}\ddot{v}\right)}{2\left(f\dot{v}^2+2\dot{\mathtt R}\dot{v}\right)^{\frac{5}{2}}},~~~~~~~
{\mathcal K}^{+\theta}_{\theta}&=&{\mathcal K}^{+\phi}_{\phi}=\frac{f\dot{v}+\dot{\mathtt R}}{{\mathtt R}\sqrt{f\dot{v}^2+2\dot{\mathtt R}\dot{v}}},
\eea
where we note that $\partial\theta/\partial\tau=\partial\varphi/\partial\tau=0$. To compute the components of extrinsic curvature for $\Sigma^-$ we begin with the connection for the interior spacetime. Using Eqs.~(\ref{gencon}) and (\ref{finaleqcontor}) this connection reads
\be\label{connectiongen}
\tilde{\Gamma}^\mu_{\,\alpha\beta}=\Gamma^\mu_{\,\alpha\beta}-\f{\kappa^2}{4\phi}\left[{S}_\beta^{\,\,\,\mu}{u}_\alpha+{S}_\alpha^{\,\,\mu}{u}_\beta+{S}_{\alpha\beta}{u}^\mu\right]+\f{1}{2\phi}\left[\delta_\beta^{\,\,\mu}\nabla_\alpha\phi-{g}_{\alpha\beta}\nabla^{\mu}\phi\right].
\ee
As we mentioned earlier, for a randomly oriented spin source $\langle {S}^{\mu\nu}\rangle=0$, hence, Eq.~(\ref{EXC}) can be rewritten as follows
\bea\label{EXCr}
{\mathcal K}^{-}_{ab}=-n_{\mu}^{-}\Bigg[\frac{\partial^2x_{-}^{\mu}}{\partial y^a\partial y^b}+\Gamma^{\mu-}_{\nu\sigma}\frac{\partial x_{-}^{\nu}}{\partial y^a}\f{\partial x_{-}^{\sigma}}{\partial y^b}+\f{1}{2\phi}\left(\delta_\sigma^{\,\,\mu}\nabla_\nu\phi-{g}_{\nu\sigma}\nabla^{\mu}\phi\right)\frac{\partial x_{-}^{\nu}}{\partial y^a}\f{\partial x_{-}^{\sigma}}{\partial y^b}\Bigg].
\eea
A straightforward calculation reveals that 
\bea
{\mathcal K}^{-t}\,\!\!_t&=&0,~~~{\mathcal K}^{-\theta}\,\!\!_\theta={\mathcal K}^{-\varphi}\,\!\!_\varphi=\f{\sqrt{1-r_\Sigma^2}}{r_{\Sigma}a(t)}.
\eea
It is not difficult to show that the above relations are the only non-vanishing components of extrinsic curvature of $\Sigma^-$. For example, bearing in mind that $\phi=\phi(t)$ along with considering nonzero components of $\Gamma^{\mu-}_{\nu\sigma}$ we get
\be\label{otherKs}
{\mathcal K}^{-}_{t\theta}={\mathcal K}^{-}_{t\varphi}={\mathcal K}^{-}_{\theta\varphi}=0.
\ee
The jump for the normal derivative of the BD scalar field across $\Sigma$ is found as
\bea\label{jumpbdsf}
\left[\phi_{,n}\right]&=&n^{\!\!+a}\phi^+_{,a}-n^{\!\!-a}\phi^-_{,a}=n^{\!\!+v}\phi^+_{,v}+n^{\!\!+{\mathtt R}}\phi^+_{,{\mathtt R}}-n^{\!\!-r}\phi^-_{,r}\nn
&=&\f{1}{\left(f\dot{v}^2+2\dot{{\mathtt R}}\dot{v}\right)^{\f{1}{2}}}\left\{(\dot{{\mathtt R}}+f\dot{v})\phi^+_{,{\mathtt R}}-\dot{v}\phi^+_{,v}\right\},
\eea
where use has been made of the contravariant components of the normal vector fields
\be\label{contnorms}
n^{\!\!-\mu}=[0,\f{\sqrt{1-r^2}}{a(t)},0,0],~~~~~~~n^{\!\!+\mu}=\f{1}{\left(f\dot{v}^2+2\dot{{\mathtt R}}\dot{v}\right)^{\f{1}{2}}}[-\dot{v},\dot{{\mathtt R}}+f\dot{v},0,0].
\ee
Thus, from Eqs.~(\ref{juncmbd}) and (\ref{juncmbd1}), the jump in the normal derivative of BD scalar field and its continuity across $\Sigma$ gives
\be\label{jumpbdsff}
\f{1}{\left(f\dot{v}^2+2\dot{{\mathtt R}}\dot{v}\right)^{\f{1}{2}}}\left\{(\dot{{\mathtt R}}+f\dot{v})\phi^+_{,{\mathtt R}}-\dot{v}\phi^+_{,v}\right\}=-\f{\phi_{\Sigma}}{2\omega+3}\left\{{\cal K}^{+t}\,\!\!_t+2\left({\cal K}^{+\theta}\,\!\!_\theta-K^{-\theta}\,\!\!_\theta\right)\right\},~~~~\phi^+|_{\Sigma}=\phi^-|_{\Sigma}=\phi_\Sigma.
\ee
Once the matter distribution on boundary surface $\Sigma$ is specified, the behavior of BD scalar field in the exterior region along with the exterior metric function will be determined through Eqs.~(\ref{FFF1}), (\ref{juncmbd}) and Eq.~(\ref{jumpbdsff}). If we assume the boundary surface to be free of energy density and pressure, i.e., $\rho_{\rm s}=0$ and $p_{\rm s}=0$, we find from Eq.~(\ref{juncmbd}) ${\cal K}^{+\theta}\,\!\!_\theta={\cal K}^{-\theta}\,\!\!_\theta$ and ${\cal K}^{+t}\,\!\!_t=0$. Therefore, the continuity of $(\theta,\theta)$ and $(t,t)$ components of extrinsic curvature across $\Sigma$ leads to the following relations 
\bea\label{extquant}
f\dot{v}+\dot{{\tt R}}&=&\sqrt{1-r_\Sigma^2},\label{aa2z}\\
\dot{v}^2\left[(ff_{,{\tt R}}+f_{,v})\dot{v}+3f_{,{\tt R}}\dot{{\tt R}}\right]&+&2\left(\dot{v}\ddot{{\tt R}}-\dot{{\tt R}}\ddot{v}\right)=0,\label{aa1z}
\eea
where we have used the relations given in Eq.~(\ref{FFF1}). Taking derivatives of Eq.~(\ref{aa2z}) and the first part of Eq.~(\ref{FFF1}) we obtain
\be\label{A17n}
2\dot{{\tt R}}\ddot{v}-\dot{f}\dot{v}^2=0,~~~~~~2\dot{{\tt R}}\ddot{{\tt R}}+\dot{f}\dot{v}\left(2\dot{{\tt R}}+f\dot{v}\right)=0,
\ee
whereby we get
\be\label{RECn}
\ddot{{\tt R}}\dot{v}=-\ddot{v}\left(2\dot{{\tt R}}+f\dot{v}\right).
\ee
Substituting for $\ddot{{\tt R}}\dot{v}$ and $\dot{{\tt R}}\ddot{v}$ from the above relations into Eq.~(\ref{aa1z}) and simplifying the result we finally get
\be\label{KTTPFn}
{\cal K}^{+t}\,\!\!_t=-\frac{f_{,v}\dot{v}^2}{2\dot{{\tt R}}}=0,
\ee
which implies that $f(v,{\tt R})$ must be a function of ${\tt R}$, only. Now, solving Eq.~(\ref{aa2z}) and the first part of Eq.~(\ref{FFF1}) we find the four-velocity of the boundary as
\be\label{4Vn}
{\rm V}^{\alpha}=\left(\dot{v},\dot{\tt R},0,0\right)=\left[\f{\sqrt{1-r_\Sigma^2}\mp\sqrt{1-f-r_\Sigma^2}}{f},\pm\sqrt{1-f-r_\Sigma^2},0,0\right],
\ee
where negative (positive) sign denotes contracting (expanding) regime. The second part of Eq.~(\ref{jumpbdsff}) implies that the BD scalar field must be homogeneous in the exterior region thus, $\phi^{+}_{,{\tt R}}=0$. Using this result in the first part of Eq.~(\ref{jumpbdsff}) we find $\phi^{+}_{,v}=0$ implying that the BD scalar field must be constant in the exterior region. It is therefore seen that the exterior region is a static spacetime with dynamical boundary. From the second part of Eq.~(\ref{FFF1}) and radial component of the four-vector velocity we get $r_{\Sigma}^2\dot{a}^2=1-f-r_{\Sigma}^2$. From the exterior viewpoint the apparent horizon is located at $f=0$ and it would meet the boundary surface if the collapse velocity satisfies
\be\label{horrr21}
r_{\Sigma}=\f{1}{\sqrt{1+\dot{a}^2}}.
\ee
The above expression is nothing but the condition of apparent horizon formation in the interior spacetime, i.e., Eq.~(\ref{rah}) taking ${\cal R}_{\rm ah}=r_\Sigma a$. We therefore conclude that in the singular case where the collapse velocity diverges, the initial radius of the boundary surface must be set to zero in order to avoid horizon formation. This situation is obviously nonphysical as the collapsing body has nonzero initial radius and consequently there is no minimum value for $r_{\Sigma}$ so that the apparent horizon can be avoided. On the other hand, if the velocity of collapse scenario is bounded, as it happens in the nonsingular case, the boundary of the collapsing object can be chosen so that the formation of apparent horizon is avoided, see section (\ref{DAH}). Thus, the bounce event can be visible to the observers in exterior spacetime. 
\section{Concluding Remarks}\label{summconc}
The longstanding issue of gravitational collapse of compact objects in the Universe has been an important subject of astronomical study. Over the past decades, much efforts have been devoted to understanding the dynamics and final fate of a collapse process. The models that have been proposed so far describe both singular and nonsingular collapse scenarios, where in the former the collapse process necessarily ends in formation of a spacetime singularity, either naked or covered by black hole horizon, and, in the latter the singularity event is prevented. Several research works in line of singularity avoidance have been carried out in both classical and quantum collapse scenarios. The results show that modifications to GR in classical regime or quantum corrections can alter the whole collapse dynamics and finally replace the spacetime singularity that occurs in GR by a nonsingular bounce~\cite{frbamba,ECNON,JaHaZia,QuanGra,QuanGra1,joshicoll,Husain,malafa,otherqgc,nonsinspin}. In the present work we tried to deal with this issue in a modified version of BD theory where the spin effects of collapsing matter are introduced as additional degree of freedom. In the original version of this theory it is shown that the collapse process leads to singularity formation covered by an event horizon~\cite{saulbhd}. However, when spin effects are present, the dynamical behavior of the collapse setting would be completely different. The study of modified field equations showed that the spin effects come into play as a repulsive pressure that oppose against the attractive pull of gravity and finally replace the spacetime singularity by a nonsingular event. Moreover, the spin effects can change horizon dynamics so that in the singular case where these effects are absent the radius of apparent horizon decreases to finally cover the singular region. While, in the nonsingular case there is a minimum value for the apparent horizon curve so that a suitable choose of the initial radius of the collapsing body could prevent horizon formation. The exterior solution was studied through a smooth matching of interior to exterior spacetimes using Israel-Darmois junction conditions. It was found the exterior spacetime is described by a Schwarzschild metric in retarded null coordinates which admits a dynamical boundary. We therefore concluded that the apparent horizon does not meet the boundary through which the interior and exterior regions are joined, hence, the bounce event can be detected by external observers.
\par\vspace{0.2cm}
{\bf Data Availability Statement} Data sets generated during the current study are available from the corresponding author on reasonable request.

\end{document}